\documentclass[11pt,a4paper]{article}
\usepackage{jcappub}
\usepackage[utf8]{inputenc}
\pdfoutput=1 

\usepackage{subcaption}
\usepackage{graphicx}%
\usepackage{multirow}%
\usepackage{amsmath,amssymb,amsfonts}%
\usepackage{amsthm}%
\usepackage{mathrsfs}%
\usepackage[title]{appendix}%
\usepackage{xcolor}%
\usepackage{textcomp}%
\usepackage{manyfoot}%
\usepackage{booktabs}%
\usepackage{algorithm}%
\usepackage{algorithmicx}%
\usepackage{algpseudocode}%
\usepackage{listings}%
\usepackage{comment}
\usepackage[normalem]{ulem}
\usepackage[utf8]{inputenc}
\usepackage{lmodern}         % recommended
\usepackage{fontawesome}
\usepackage{xcolor}

\newcommand{\be}{\begin{equation}}
\newcommand{\ee}{\end{equation}}
\newcommand{\beq}{\begin{equation*}}
\newcommand{\eeq}{\end{equation*}}
\newcommand{\bea}{\begin{eqnarray}}
\newcommand{\eea}{\end{eqnarray}}

\newcommand{\bn}{{\mathbf n}}

\newcommand{\bv}{{\mathbf v}}

\newcommand{\HH}{{\cal H}}

\newcommand{\De}{\Delta}

\newcommand{\Omm}{\Omega_{{\rm m}}}
\newcommand{\Ommo}{\Omega_{{\rm m, 0}}}

\newcommand{\Om}{\Omega}

\newcommand{\cd}{\cdot}

\begin{document}

\title{Reconstructing the kinematics of Laniakea using Type Ia Supernovae}

\author[a]{Francesco Sorrenti,}
\author[b]{Erick Pastén,}
\author[c]{Leonardo Giani }

% The "\note" macro will give a warning: "Ignoring empty anchor..."
% you can safely ignore it.

\affiliation[a]{Institut de Ciències del Cosmos, Universitat de Barcelona (ICCUB), Martí i Franquès, 1, 08028 Barcelona, Spain}
\affiliation[b]{Departamento de F\'{i}sica, Universidad de Santiago de Chile, Avenida V\'{i}ctor Jara 3493, Estaci\'{o}n Central, 9170124, Santiago, Chile}
\affiliation[c]{Swinburne University of Technology, Hawthorn VIC 3122, Australia}

% e-mail addresses: one for each author, in the same order as the authors
\emailAdd{francescosorrenti@icc.ub.edu}
\emailAdd{erick.contreras@usm.cl}
\emailAdd{lgiani@swin.edu.au}

\abstract{
%At low redshift, anisotropies in the luminosity distance are commonly modelled through a multipolar expansion attributed to peculiar velocities. 
We develop a kinematic framework that relates the monopole, dipole, and quadrupole of the luminosity distance to an ellipsoidal peculiar velocity field describing the dynamics of the Laniakea supercluster. By properly accounting for the transformations between the CMB and Laniakea reference frames and selecting Type Ia supernovae within the volume associated with the superstructure, we show that luminosity distance multipoles encode the expansion and shear of the local velocity field. This allows the ellipsoidal kinematics of Laniakea to be inferred directly from supernova observations. 

Our results provide a physically motivated interpretation of luminosity distance anisotropies within the volume dynamically dominated by Laniakea as signatures of its large-scale kinematics. The framework developed here suggests that future supernova observations can enable the independent reconstruction of large-scale structures in the local Universe
}


\maketitle

%%%%%%%%%%%%%%%%% BODY OF PAPER %%%%%%%%%%%%%%%%%%

\section{Introduction}

The Cosmological Principle is a cornerstone of modern cosmology, asserting that the Universe is homogeneous and isotropic on sufficiently large scales. Under this assumption, any anisotropy inferred from low-redshift observables---such as the Hubble constant $H_0$, the deceleration parameter $q_0$, or the matter density $\Ommo$---should originate from local large-scale structure features. Supernova surveys provide an ideal testing ground for these ideas. Numerous works have attempted to quantify anisotropies in the luminosity distance by fitting dipole \cite{Bonvin:2006en,Javanmardi_2015,Colin_2019,Wang_2017,Bengaly:2017slg,sorrenti2023dipolepantheonsh0esdata} or higher multipole corrections \cite{Dhawan_2022,Sah_2025,Sorrenti2024ztg} to different supernovae compilations. These anisotropic signatures are phenomenologically attributed to peculiar motions with respect to the cosmic rest frame, reflecting the structure of the nearby velocity field. However, the mapping between the fitted multipoles and the underlying kinematics of the local velocity field, such as its expansion, shear, and the observer’s displacement from the dynamical center, is highly nontrivial and hard to model with standard perturbed FLRW machinery. As a result, the inferred amplitudes cannot be straightforwardly interpreted in terms of the geometry and dynamics of the local flow.

Environmental contributions to our cosmological inference are particularly important to understand. Indeed, recent analyses have identified anomalously large Bulk Flows (BF) in the local Universe with respect to the $\Lambda$CDM predictions~\cite{Watkins2026,Whitford_2023,Watkins_2023}. Furthermore, it has been suggested that local inhomogeneities could be responsible for some of the recent challenges faced by the standard cosmological model, in particular the Hubble tension, the evidence for dynamical Dark Energy, and the challenges to the Cosmological Principle~\cite{SylosLabini:2026gdf,Macpherson:2026bzj,Giani_2025b,Giani_2025a,Aluri_2023, Pasten:2023rpc, Luongo:2025kfs, Abghari:2024eja,Ellis:1984, Blake:2002kx, Tiwari:2016, Colin:2017juj,Secrest:2020has,Siewert:2021, Secrest:2022uvx, Dalang:2021ruy,Guandalin:2022tyl,Cheng:2023eba,daSilveiraFerreira:2024ddn}. In light of these results, several studies have considered the velocity field as a probe of the Cosmological Principle and of relativistic effects beyond Newtonian perturbation theory~\cite{Galoppo:2026rin,Giani:2026soh,Kalbouneh:2022tfw,Heinesen:2023lig}.  There has also been an increasing interest in using peculiar velocities and the velocity power spectrum as observables to constrain cosmological parameters~\cite{macaulay,6df_pec, Howlett:2017asq, Piras:2025gnx, Carreres:2023nmf, Koda:2013eya}.  

On the other hand, reconstructing the local peculiar velocity field remains a challenging task. At low redshift, the degeneracy between cosmological expansion parameters and local kinematic effects complicates any attempt to disentangle global and environmental contributions without imposing strong modeling assumptions. In practice, commonly adopted approaches—such as reconstructions based on density survey data \citep{Carrick_2015}—rely on Newtonian relations between the density contrast and peculiar velocities. These methods generally neglect vorticity, assume potential flows, and may be sensitive to biases induced by local under- and over-densities or incomplete sky coverage. Recent analyses of Pantheon$+$ supernovae~\cite{Sorrenti_2024} have nevertheless reported evidence for a coherent local infall pattern, consistent with earlier Newtonian reconstructions analysis of the nearby velocity field \citep{Pasten_2024} and with the study by Giani et al.\ \citep{Giani:2023aor} using CosmicFlows-4 data, who characterised an effective model of the supercluster \emph{Laniakea}. Laniakea is one of the largest structures in the nearby Universe~\cite{Tully:2014gfa}, encompassing the Milky Way and thousands of galaxies over scales of several hundred megaparsecs. In \cite{Giani:2023aor}, following the theoretical investigation of \citep{Giani:2021gbs}, the velocity field within the Laniakea supercluster was shown to be well described by an anisotropic ellipsoidal model, characterised by non-zero shear and a negative volume expansion. Interestingly, the inferred sign of the expansion reverses when spherical symmetry is imposed, highlighting the importance of anisotropic kinematics in the study of the local flow.

Motivated by these considerations, the present work builds upon the analysis of \cite{Sorrenti2024ztg}, where the monopole, dipole, and quadrupole of the luminosity distance were directly fitted using Type~Ia supernovae (SNe Ia), to introduce a physically motivated model for the local peculiar velocity field of the Laniakea supercluster, described as an anisotropic ellipsoidal flow with a constant value for expansion and shear \cite{Giani:2023aor}. We show that the observed multipoles in the luminosity distance are consistently related to the expansion, coherent motion, and shear of the local flow. This allows us to reinterpret the luminosity distance anisotropies inside the Laniakea volume in terms of the underlying ellipsoidal kinematics. Beyond its application to Laniakea, this approach therefore opens the possibility of using Type Ia supernovae as direct kinematic tracers of large-scale structure. Currently, the large-scale structure (LSS) matter distribution is primarily mapped through galaxy density fields. The present work is therefore the first application of an independent method that can be applied in the future to other structures, once sufficient supernovae observations become available (in particular from LSST \citep{Ivezi__2019} and ZTF \citep{Rigault:2024kzb}). 

The structure of the paper is as follows. In Sec.~\ref{sec:theory}, we introduce the kinematic description of the Laniakea velocity field. In particular, we show how the perturbed luminosity distance and the multipolar decomposition of the corrected redshift can be interpreted in terms of the expansion and shear of the local peculiar velocity field. In Sec.~\ref{sec:data_methodology}, we introduce the Pantheon+ dataset and describe the methodology used to constrain the kinematic parameters inside the Laniakea region. In Sec.~\ref{sec:results}, we discuss the resulting velocity field and compare it with independent reconstructions of the local flow such as the one made in \cite{Giani:2023aor}. In Sec.~\ref{sec:mock_validation}, we validate our methodology against mock datasets. Finally, Sec.~\ref{sec:conclusions} summarizes our main conclusions.

\vspace{0.2 cm}
\textbf{Notation:} We considered a spatially flat FLRW Universe where the line element is given by
\be
ds^2 = a^2(t) [-(1+2\Psi)c^2 dt^2 + (1-2\Phi)\delta_{ij}dx^i dx^j] \,,
\ee
in the conformal Newtonian gauge. Here, $c$ is the speed of light and $(\Phi, \Psi)$ are the Bardeen potentials. We assumed the Einstein summation convention, and tensors are written in boldface. An overdot denotes a derivative with respect to conformal time $t$. Consequently, $\HH=\dot{a}/a$ is the comoving Hubble parameter, and $H=\dot{a}/a^2$ is the physical Hubble parameter. The scale factor at the present time, $a_0=a(t_0)$, is normalized to unity. Finally, $r(z)$ denotes the comoving distance at redshift $z$.
\section{Theoretical model}\label{sec:theory}

The analysis presented in \cite{Giani:2023aor} is performed in a reference frame comoving with the Laniakea supercluster, where the bulk velocity of the structure is subtracted from the velocity field. In contrast, the multipolar parameters inferred in \cite{Sorrenti2024ztg} are extracted from the point of view of an observer located in the Solar System. In order to consistently relate both approaches, it is therefore necessary to explicitly introduce and connect the corresponding reference frames.

\subsection{The Laniakea supercluster}

Catalogues of galaxy peculiar velocities, obtained by removing the Hubble flow from the redshifts of galaxies with known distances, can be used to reconstruct the density and velocity fields of the dark matter fluid in the local Universe. In particular, the latest reconstructions \cite{Courtois:2022mxo} of the velocity and density fields from the Cosmic-Flows 4 (CF4) catalogue \cite{Tully_2023} (and its updated version CF4++ \cite{Courtois:2025xcs}) have reached an impressive depth of $\approx 500\; \mathrm{Mpc}/h$, and have been recently used to estimate how our cosmic neighbourhood can impact cosmological inference~\cite{Galoppo:2026rin,Galoppo:2025hzy}.

The reconstructed velocity field can be used to partition the local Universe {\color{blue}into} gravitational basins of attraction, as done for example in~\cite{Dupuy:2019blz,Dupuy:2023ffz,Valade_2024}. These basins can be defined by analyzing the streamlines of the velocity field, which are curves tangent to it at any point on a fixed spatial hypersurface. To build the streamlines from the reconstructed velocity field $\vec{v}$, one starts from an arbitrary initial point on the grid, labelled \textit{seed}, and integrates $\vec{v}$ between neighboring cells. The curve connecting the two tangent vectors at these points defines the beginning of the streamline. One then repeats the process iteratively until a critical point is reached, i.e., an attractor, where the streamline stops. The spatial volume containing streamlines converging towards the same attractor defines a basin of attraction, representing a region of space where the trajectories of test particles are bounded.

The Laniakea supercluster, containing our own Milky Way, was first defined in Ref.~\cite{Tully:2014gfa} using the velocity reconstruction obtained from Cosmicflows-2 (CF2)~\cite{Tully:2013wqa}, and its boundaries have been later revisited in~\cite{Valade_2024,Courtois:2025xcs} using improved techniques and better data. Inspired by the Bianchi IX gravitational collapse of matter inhomogeneities developed in \cite{Giani:2021gbs}, the authors of \cite{Giani:2023aor} proposed an effective modeling of the velocity field within Laniakea using a triaxial model. The structure is approximated as {\color{blue}a} homogeneous ellipsoid, and the velocity field is described by an anisotropic expansion of its volume. In this work, we use the ellipsoidal volume defined in \cite{Giani:2023aor} to identify the boundary of Laniakea, and independently determine its local kinematics by developing a suitable formalism in Sec.~\ref{subsec:kinematic_structure}.

\subsection{The kinematic structure of Laniakea}\label{subsec:kinematic_structure}

We define three reference frames throughout this analysis: 
\begin{itemize}
\item the vector $\mathbf{x}_{\rm src/L}$ denotes the position of a source as measured by an observer comoving with Laniakea and located at its dynamical center defined by the model of \cite{Giani:2023aor}; 
\item the vector $\mathbf{x}_{\rm src/\odot}$ corresponds to the position of the same source as measured from the Solar System. The Solar System is displaced from the centre of Laniakea by a vector $\mathbf{p}$ measured in the Laniakea frame, and the observational frame, which could be equatorial or supergalactic, is rotated with respect to the principal axes of the Laniakea ellipsoid; 
\item the background comoving frame, at rest with respect to the CMB, denoted by $\mathbf{x}_{\rm src/CMB}$.
\end{itemize}
If $\mathbf{R}$ denotes the rotation matrix connecting the Solar System frame to the frame defined by the Laniakea principal axes, we can write:
\begin{eqnarray}
    \mathbf{x}_{\rm src/L}=\mathbf{R}\cdot\mathbf{x}_{\rm src/\odot}+\mathbf{p}\,.
    \label{positions}
\end{eqnarray}
Assuming $\mathbf{R}$ to be constant over the timescales relevant to this analysis, the corresponding velocities transform as:
\begin{eqnarray}
    \mathbf{v}_{\rm src/L}=\mathbf{R}\cdot \mathbf{v}_{\rm src/\odot}+\mathbf{v}_{\rm \odot/L}\,,
\end{eqnarray}
where $\mathbf{v}_{\rm \odot/L}$ is the velocity of the Sun with respect to the Laniakea center. Following \cite{Giani:2023aor}, the velocity field in the Laniakea frame can be written as
\begin{eqnarray} \label{eq:gradient_laniakea}
    \mathbf{v}_{\rm src/L}=\mathbf{\Delta H} \cdot \mathbf{x}_{\rm src/L}\,,
\end{eqnarray}
modelling it as an ellipsoidal flow characterised by three parameters collected in a diagonal tensor $\mathbf{\Delta H}=\mathrm{diag}(\Delta \rm{H}_a,\Delta \rm{H}_b,\Delta \rm{H}_c)$. It is possible to show (see Appendix \ref{app0} for more details) that the same velocity in the CMB rest frame can be described as 
\begin{eqnarray}
    \mathbf{v}_{\rm src/CMB}=A(\bar{z})\,\left[\mathbf{E}\cdot\mathbf{x}_{\rm src/\odot}+\mathbf{d}\right]%\left[ 1-\mathcal{H}(\bar{z})r(\bar{z})\right]\,,
    \label{peculiarvelocityfield}
\end{eqnarray}
where we have also introduced the dependence on the redshift $\bar{z}$ corrected for our own motion with respect to the CMB, with $A(\bar{z})$ given by Eq.~\eqref{theory:prefactor} and 
\begin{eqnarray}
    \mathbf{E}&=&\mathbf{R}^{-1}\cdot \mathbf{\Delta H} \cdot \mathbf{R},\label{eq:similarity}\\
    \mathbf{d}&=&\mathbf{R}^{-1}\cdot (\mathbf{\Delta H} \cdot \mathbf{p}- \mathbf{v}_{\rm \odot/L})+\mathbf{v}_{\rm \odot/CMB}\,.
\end{eqnarray}
Here, $\mathbf{v}_{\rm \odot/CMB}$ is the velocity of the Solar System with respect to the CMB, and $\mathbf{d}$ is the dipolar component of the velocity field measured from the CMB rest frame, which can be interpreted as the bulk motion velocity. Following \citep{Giani:2023aor}, the ellipsoidal flow is assumed to be irrotational. This is reasonable since in the standard model vorticity is sourced only by non-linear dynamics at small scales, and it is completely negligible on scales of order\,$\sim100\,\mathrm{Mpc}/h$, which are our scales of interest. As a consequence, the antisymmetric part of the tensor $E_{ij}$ effectively cancels out. This allows the tensor to be decomposed into a scalar and a traceless component,
\begin{eqnarray} \label{eq:deco}
E_{ij} = \frac{1}{3}\theta\,\delta_{ij} + \sigma_{ij}\,,
\end{eqnarray}
where $\theta = \partial_i v_i$ is the scalar expansion encoding the isotropic part of the flow, and 
$\sigma_{ij}$ is the shear tensor describing the anisotropic expansion. Consequently, in our ellipsoidal model
\begin{eqnarray}
\theta &=& {\rm Tr}(\mathbf E)\,,\label{eq:theta}\\
\sigma_{ij}&=&E_{ij}-\frac13\theta\delta_{ij}\,.
\label{eq:sigma}
\end{eqnarray}

Since $\mathbf E$ and $\mathbf{\Delta H}$ are related by a similarity transformation,
their scalar invariants do not change.
Therefore, the scalar expansion and the shear eigenvalues inferred in the CMB rest-frame are identical to the trace of the ellipsoidal expansion tensor and the eigenvalues of its traceless part in the Laniakea frame.

From Eq.~\eqref{eq:gradient_laniakea}, we can see that the gradient of the velocity field in the Laniakea frame is described by $\mathbf{\Delta H}$. The eigenvectors of this tensor in the Laniakea frame are simply the orthonormal basis vectors
\begin{equation}
\hat{\mathbf u}_a=(1,0,0)\,,
\qquad
\hat{\mathbf u}_b=(0,1,0)\,,
\qquad
\hat{\mathbf u}_c=(0,0,1)\,.
\end{equation}
From the similarity relation~\eqref{eq:similarity}, we also have that the eigenvectors of $\mathbf{E}$ are 
\begin{equation}
\label{eq:eigen}
\hat{\mathbf e}_i
=
\mathbf R^{-1} \cdot 
\hat{\mathbf u}_i\,.
\end{equation}
This means that the eigenvectors of the ellipsoidal velocity field component measured from the Solar System correspond to the principal axes of the Laniakea flow expressed in the Solar System itself. Further information about the characteristics of the eigenvectors with respect to the ellipsoidal Laniakea geometry can be found in Appendix~\ref{appA}. 

Within the formalism developed here, we can identify the key kinematic parameters inferred in \cite{Giani:2023aor} which we aim to recover with our method. The effective isotropic expansion associated with the Laniakea velocity field, as inferred from the best model over the CosmicFlows-4 sample \cite{Tully_2023}, is estimated to be
\begin{eqnarray}
    \theta_L = 3\Delta \rm{H}_{\rm tot} \simeq -3.33 \pm 0.27 .
\end{eqnarray}
The corresponding shear components can be obtained from the eigenvalues of the
velocity gradient tensor describing the ellipsoidal flow. In the principal axis
frame of Laniakea these are given by
\begin{eqnarray}
    \sigma^L_{a}&=&\Delta  \rm H_a -\Delta  \rm{H}_{\rm tot} \simeq -2.29\pm 0.70\,, \\
    \sigma^L_{b}&=&\Delta  \rm H_b -\Delta  \rm \rm{H}_{\rm tot} \simeq 0.46 \pm 0.55\,, \\
    \sigma^L_{c}&=&  \rm \Delta H_c -\Delta  \rm \rm{H}_{\rm tot} \simeq 1.80 \pm 0.45\,,
\end{eqnarray}
where $3 \Delta \rm{H}_{\rm tot}=\Delta \rm{H}_a+\Delta \rm{H}_b+\Delta \rm{H}_c $. A summary of these parameters and other features of the Laniakea ellipsoidal model are summarized in Tab.~\ref{Tab:Lparameters}.

\begin{table}[!htbp]
\centering

\renewcommand{\arraystretch}{1.35}

\begin{tabular}{|l|c|}
\hline\hline

Average density contrast $\langle\delta\rangle$
& $0.012$ \\ [1.5mm]

\hline
Spherical radius scale $R_{\rm eq}$
& $110$ \,\footnotesize{Mpc/$h$}\\[1.5mm]

\hline

Center (SGX,SGY,SGZ)
& $(-9.23,\,-47.06,\,85.08)$  \footnotesize{Mpc/$h$}\\ [1.5mm]

\hline
Semi-axis $a$ length
& $48.61$ \footnotesize{Mpc/$h$} \\

Semi-axis $b$ length
& $128.74$ \footnotesize{Mpc/$h$} \\

Semi-axis $c$ length
& $166.38$ \footnotesize{Mpc/$h$} \\[1.5mm]

\hline

Semi-axis $a$ direction
& $(222.66^\circ,\,45.52^\circ)$ \\

Semi-axis $b$ direction
& $(18.88^\circ,\,41.95^\circ)$ \\

Semi-axis $c$ direction
& $(120.02^\circ,\,12.13^\circ)$ \\[1.5mm]

\hline

Expansion $\theta_L$
& $-3.33 \pm 0.27$ \footnotesize{km/s/Mpc}
\\ [1.5mm]

\hline
Shear $\sigma^{\rm L}_{a}$
& $-2.29 \pm 0.70$ \footnotesize{km/s/Mpc}\\

Shear $\sigma^{\rm L}_{b}$
& $0.46 \pm 0.55$ \footnotesize{km/s/Mpc}\\

Shear $\sigma^{\rm L}_{c}$
& $1.80 \pm 0.45$ \footnotesize{km/s/Mpc}\\[1.5mm]

\hline\hline
\end{tabular}

\caption{
Summary of the geometrical and kinematic properties of the
Laniakea supercluster reported in~\citep{Giani:2023aor}. The density and geometrical quantities are reported without error bars because they were kept fixed in the analysis of~\citep{Giani:2023aor}.
}

\label{Tab:Lparameters}

\end{table}

\subsection{The luminosity distance in terms of kinematic quantities}\label{sec:LuminosityDistance}

SNe Ia data compilations provide the distance moduli $\mu$, which are related to the model luminosity distance $d_{\rm L}$ via 
\begin{eqnarray}
\label{eq:mu_model}
\mu= 5 \log \biggl( \frac{d_{\rm L}(z, \bn)}{\rm Mpc} \biggr) + 25 \, .
\end{eqnarray}
At first order in cosmological perturbation theory, the luminosity distance of an object at observed redshift $z$ in direction $\bn$ is~\cite{Biern:2016kys, Bonvin:2005ps, Hui_2006}
\bea
\hspace{-1cm}d_\mathrm{L}(z,\bn) &=& \bar d_\mathrm{L}(z)\Bigg[1 + \frac{1}{\HH(z) r(z)}\bn\cdot(\bv_{\odot/{\rm CMB}}-\bv_{\rm src/CMB})-\frac{\bn\cdot\bv_{\rm src/CMB}}{c}  -  \Phi - \nonumber  \\
&& \left(1-\frac{1}{\HH r}\right)\left(\!\Psi+ \int_{0}^{r}\!\!dr'(\dot\Psi+\dot\Phi)\right) +\int_{0}^{r}\frac{dr'}{r} \left(1-\frac{(r-r')}{2r'}\De_\Om\right)(\Psi+\Phi) \Bigg] \,,\label{e:dLpert1}
\eea
where the background luminosity distance at low redshift in a flat $\Lambda$CDM model is
\be
\bar d_{\rm L}(z)= \frac{1+z}{H_0}\int_0^z \frac{dz'}{\sqrt{\Ommo(1+z')^3+1-\Ommo}} \,.
\ee
At low redshift, the terms $\propto 1/[\HH(z)r(z)]$ are dominant, so that we can approximate 
\begin{equation}
    d_{\rm L}(z,\mathbf{n})=\bar{d}_{\rm L}(z)\left \{1+\frac{1}{\mathcal{H}(\bar{z})r(\bar{z})}\left [\left (1- \mathcal{H}(\bar{z})r(\bar{z})\right )\mathbf{v}_{\rm src/CMB}-\mathbf{v}_{\rm \odot/CMB}\right ]\cdot \mathbf{n}\right \}\,.
    \label{eqn:PerturbeddL2}
\end{equation}
As shown in~\cite{Sorrenti2024ztg}, Eq.~(\ref{eqn:PerturbeddL2}) can be rewritten by transforming the luminosity distance corrections into redshift corrections. In other words, we can write the perturbed luminosity distance as 
\begin{eqnarray}
d_{\rm L}(z,\bn) = \bar d_{\rm L}(z_{\rm cos})(1+z_{\odot})(1+z_{\rm p})^{2}  \,,
\label{eqn:masterLuminosityDistance}
\end{eqnarray}
where we have also incorporated the redshift multiplicative corrections due to Doppler and relativistic beaming effects (see \cite{Davis2014}). We have that
\bea
\label{theory:z_sun}
z_\odot&=&\sqrt{\frac{1 + (-v_\odot)/c}{1 - (-v_\odot)/c}} - 1   \qquad   \mbox {and}\\\vspace{0.2cm}
\label{theory:z_p}
z_{\rm p}&=&\sqrt{\frac{1 + (v_{\rm p})/c}{1 - (v_{\rm p})/c}} - 1 \,,
\eea
where $v_\odot=\bn\cd\bv_{\odot/\rm CMB}$ and $v_{\rm p}$ is given by 
\begin{eqnarray}
    \label{theory:v_p1}
v_{\rm p} = \left [ 1-\mathcal{H}(\bar{z})r(\bar{z})\right ] \, (\mathbf{v}_{\rm src/CMB} \cdot \mathbf{n})
\end{eqnarray} 
According to~\cite{Sorrenti2024ztg}, it is possible to write the radial velocity as
\begin{eqnarray}
    \label{theory:v_p}
\mathbf{v}_{\rm src/CMB} \cdot \mathbf{n} = A(\bar{z})\left[\gamma+\mathbf{v}_{\rm bulk}\cdot \mathbf{n} + n^{i}n^{j}\alpha_{ij} \right] ,
\end{eqnarray} 
where $\gamma$ is the monopole, $\mathbf{v}_{\rm bulk}$ is the bulk motion, and $\alpha_{ij}$ is a traceless tensor describing the quadrupole. 
Finally, we can introduce the cosmological redshift
\begin{eqnarray}
    \label{theory:z_quadrupole}
z_{\rm cos}=\frac{1+\bar{z}}{1+z_{\rm p}} -1 \,.
\end{eqnarray}

In a cosmological context, large-scale structures are naturally described on spatial hypersurfaces of constant cosmic time \cite{Peebles:1994xt}, for which comoving coordinates provide the appropriate parametrization of positions at low redshift. Accordingly, defining $\mathbf{x}_{\rm src/\odot}=r(\bar{z})\,\mathbf{n}$, we can combine Eq.~\eqref{peculiarvelocityfield} with Eq.~\eqref{eq:deco} to define $v_{\rm p}$ as
\begin{eqnarray}
    v_{\rm p} 
    =A(\bar{z})\left [r(\bar{z})\left(\frac{1}{3}\theta\,\delta_{ij}+\sigma_{ij}\right)n^in^j+\mathbf{d}\cdot \mathbf{n}\right]\left [ 1-\mathcal{H}(\bar{z})r(\bar{z})\right ]\,.
\end{eqnarray}\label{eq:vp_final}
Comparing the equation above with Eq.~\eqref{theory:v_p}, we can express the multipolar parameters in terms of the kinematic quantities of the peculiar velocity field as
\begin{eqnarray}
    \mathbf{v}_{\rm bulk}&=&\mathbf{d}\,,\\
    \gamma&=&\frac{r(\bar{z})}{3}\theta\,,\\
    \alpha_{ij}&=&r(\bar{z})\sigma_{ij}\,.
    \label{kinematicrelations}
\end{eqnarray}

\section{Data and methodology}\label{sec:data_methodology}

We now apply the formalism developed above to SNe Ia observations in order
to constrain the kinematic parameters of the velocity flow in the region dynamically
dominated by the Laniakea supercluster. Our goal is to determine whether the anisotropies inferred from supernova luminosity
distances are consistent with the ellipsoidal velocity field describing the dynamics
of Laniakea.

We use the Pantheon+SH0ES compilation of SNe Ia \cite{Brout_2022}, including the Cepheid-calibrated SH0ES subset. This dataset~\footnote{Available at~\url{https://github.com/PantheonPlusSH0ES/DataRelease}.} provides the largest currently available sample of spectroscopically confirmed SNe Ia with uniformly calibrated distance moduli. The catalog includes the angular positions of the SNe in equatorial coordinates (ra, dec) and their redshifts, $\bar{z}$, corrected for our own motion with respect to the CMB according to the dipole measured in~\cite{Planck:2018vyg}. It also provides the associated covariance matrix, which incorporates both statistical and systematic uncertainties.

\begin{figure}[t]
\centering

\begin{subfigure}{0.47\textwidth}
    \centering
    \includegraphics[width=\textwidth]{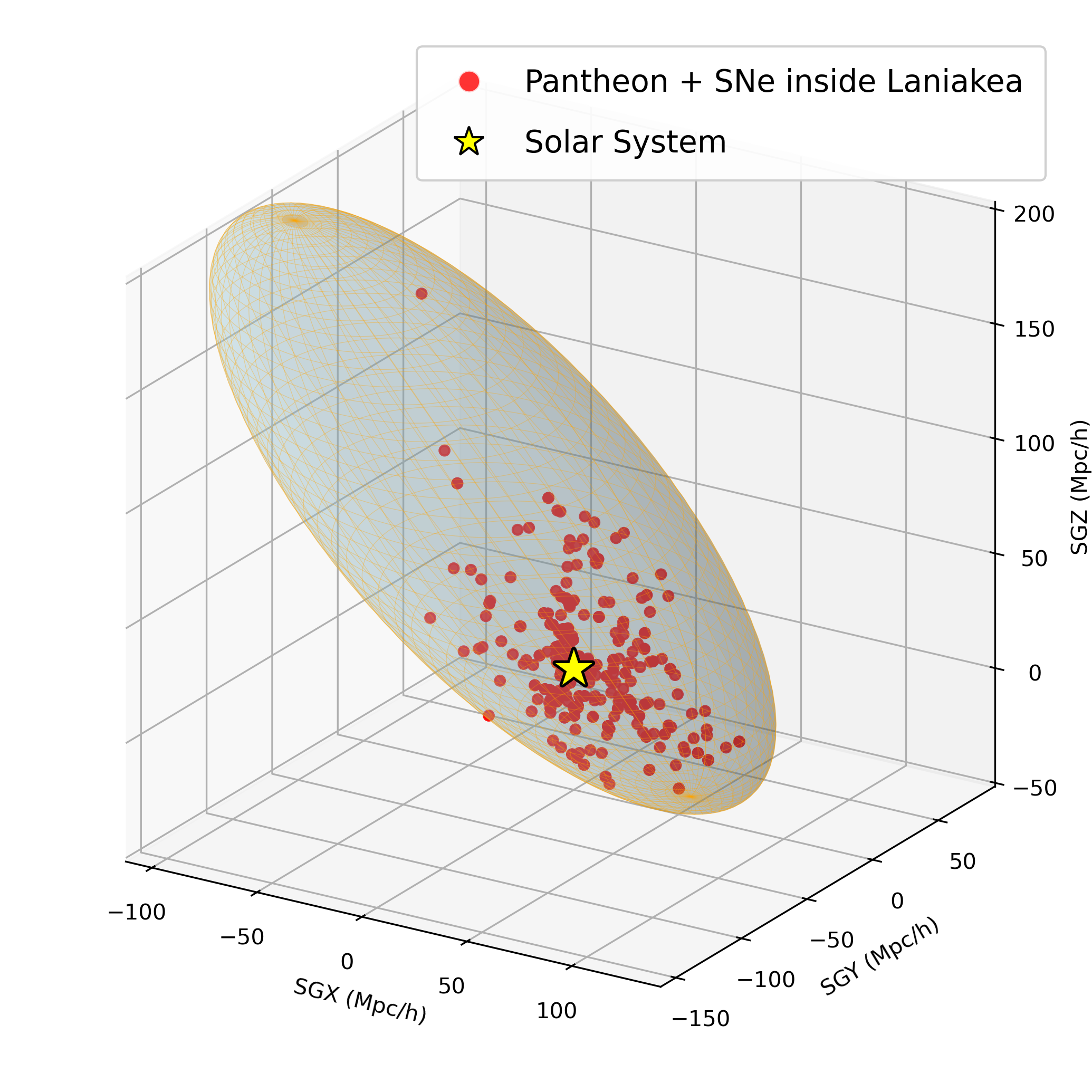}
    \label{fig:laniakea_sn_3d}
\end{subfigure}
\hfill
\begin{subfigure}{0.47\textwidth}
    \centering
    \includegraphics[width=\textwidth]{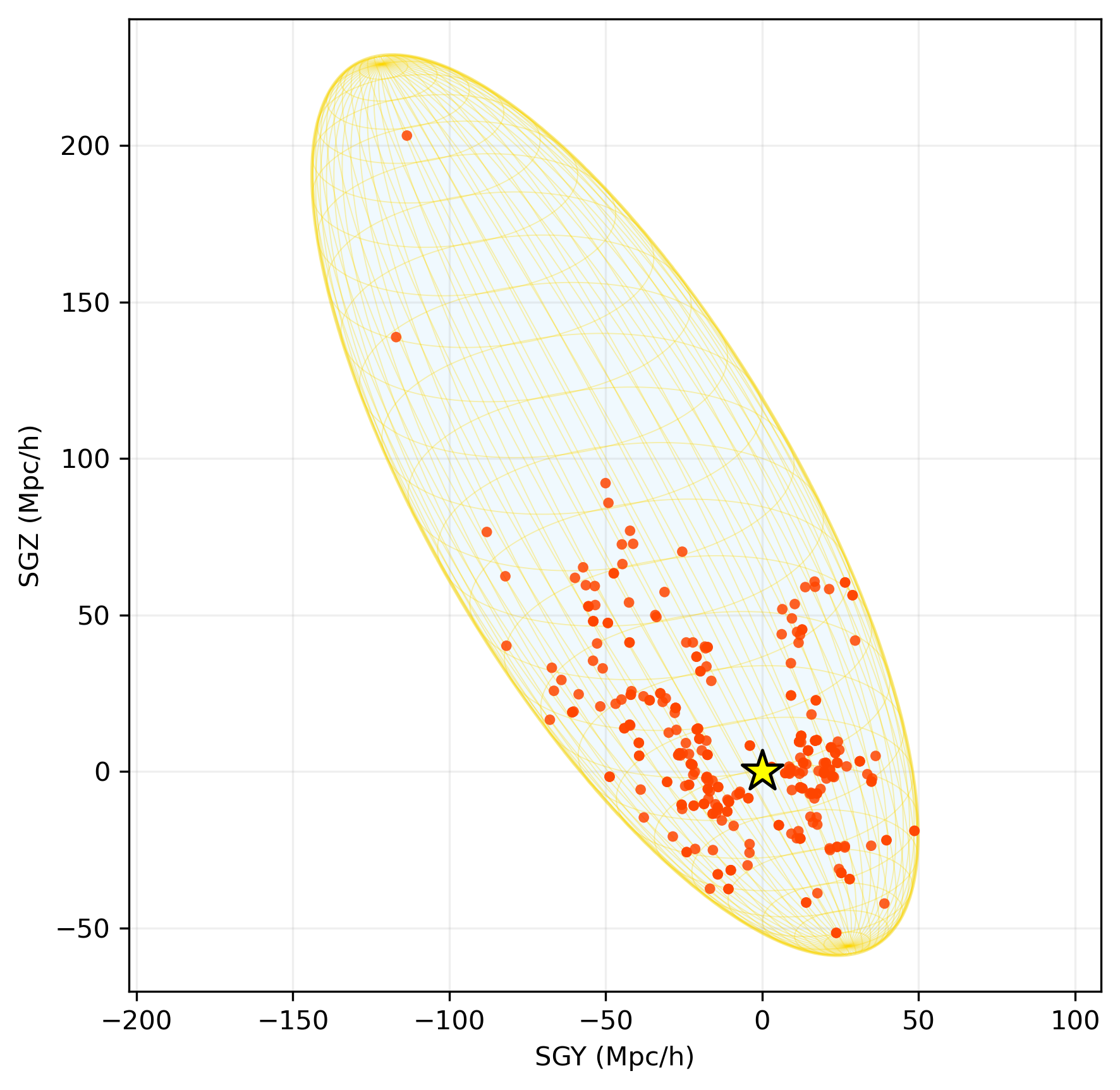}    \label{fig:laniakea_sn_sgy_sgz}
\end{subfigure}

\vspace{0.3cm}

\begin{subfigure}{0.47\textwidth}
    \centering
    \includegraphics[width=\textwidth]{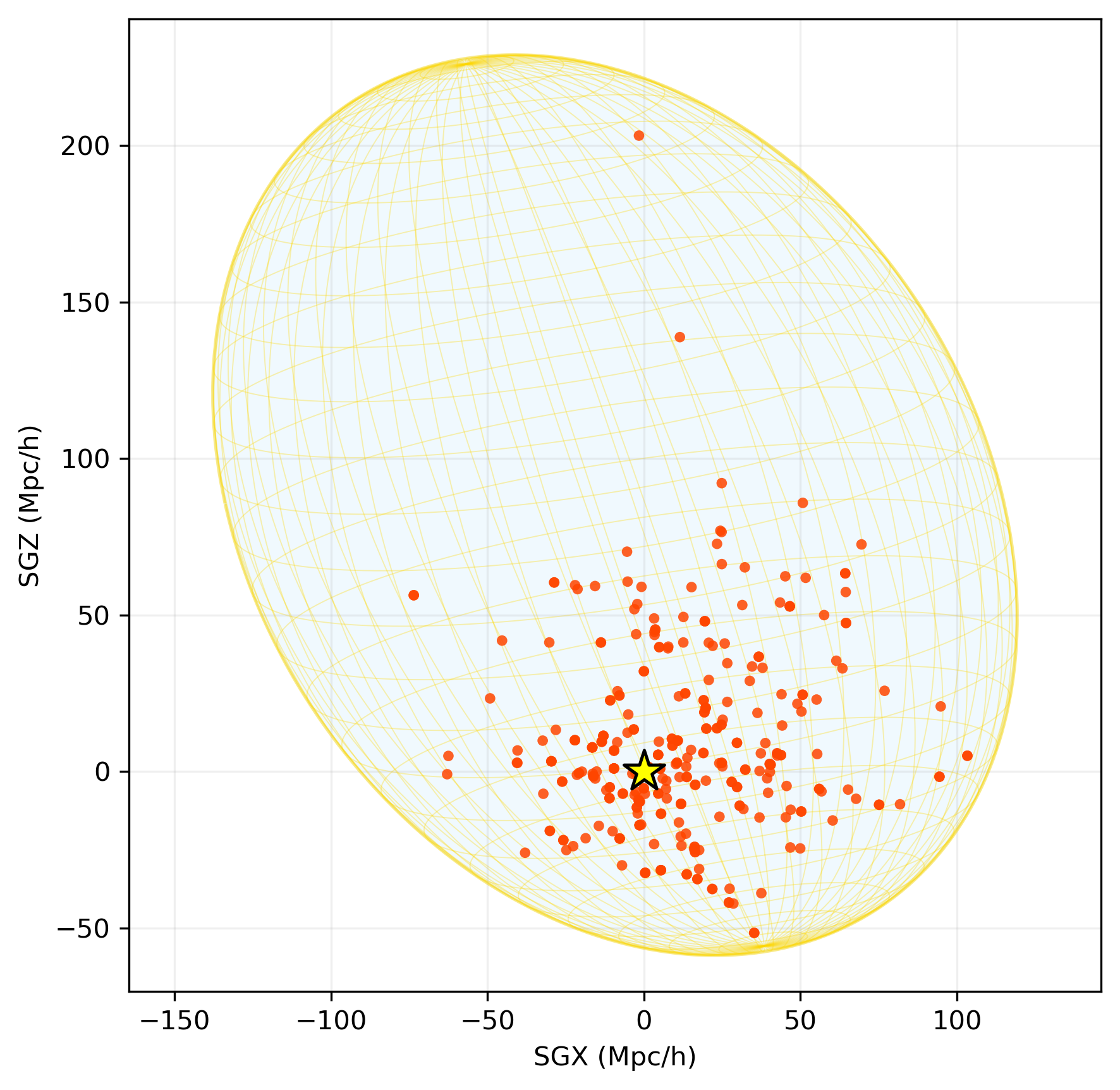}
    \label{fig:laniakea_sn_sgx_sgz}
\end{subfigure}
\hfill
\begin{subfigure}{0.47\textwidth}
    \centering
    \includegraphics[width=\textwidth]{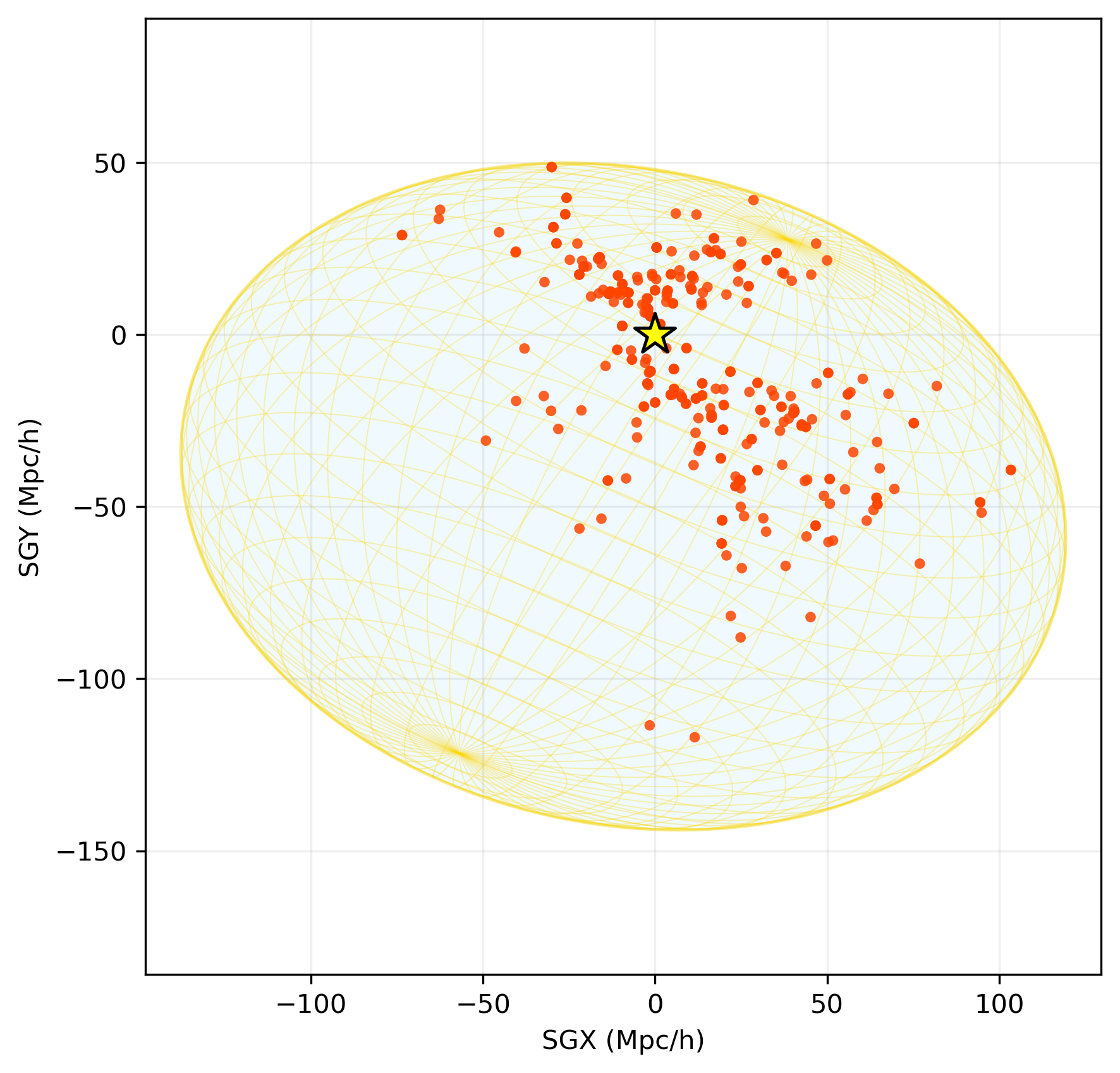}
    \label{fig:laniakea_sn_sgx_sgy}
\end{subfigure}

\caption{
Spatial distribution of Pantheon+ SNe Ia located inside the
Laniakea region in supergalactic coordinates. The top-left panel shows the
three dimensional distribution, while the remaining panels show the corresponding
two-dimensional projections onto the SGY--SGZ, SGX--SGZ, and SGX--SGY planes. The translucent ellipsoidal surface indicates the effective Laniakea volume adopted from
\cite{Giani:2023aor} while the red points represent the supernovae inside it. The yellow star marks the position of the Solar System, highlighting the off-centre location of the observer within the structure.
}
\label{fig:laniakea_sn_distribution}
\end{figure}

To assess the impact of Laniakea, the analysis is performed by separating the supernova sample into two sub-samples
according to whether their spatial position lies inside or outside the ellipsoidal volume associated with Laniakea, following the effective description of the Laniakea region presented in \cite{Giani:2023aor}. Supernovae positions are converted into Cartesian coordinates in the supergalactic frame, by assigning to each object a fiducial comoving distance inferred from its CMB-frame redshift. In practice, we use a flat $\Lambda$CDM distance with $H_0=100$\,km/s/Mpc, so that the resulting coordinates are expressed directly in $\mathrm{Mpc}/h$, matching the units of the Laniakea ellipsoidal model. At the low redshifts relevant for the Laniakea volume, this prescription reduces to $r(\bar{z})\,h \simeq c\,\bar{z}/100$, making the geometrical selection effectively independent of the absolute value of $H_0$. We further verified that deviations from the $\Ommo$ fiducial value $\Ommo=0.30$ are subdominant and negligible in our analysis. Fig.~\ref{fig:laniakea_sn_distribution} illustrates the spatial distribution
of Pantheon+ supernovae inside the Laniakea volume in supergalactic coordinates.
The ellipsoid represents the effective Laniakea boundary adopted from
\cite{Giani:2023aor}, while the position of the Solar System is indicated as
reference and highlights the off-centre location of the observer inside the ellipsoid.

We perform an MCMC exploration of the cosmological parameters $(H_0,\Ommo)$, alongside the kinematic parameter $\theta$ and the five independent elements of the matrix $\sigma_{ij}$, from which we derive the shear eigenvalues $\sigma_i$ and eigenvectors $e_i$. The cosmological parameters are constrained globally using the full supernova sample, while the kinematic parameters describing the local velocity field are divided into those inside and those outside Laniakea. In practice, this corresponds to adopting a piecewise description where Eq.~\eqref{eq:vp_final} becomes:
\begin{equation}
\label{ep:vp_cases}
v_{\rm p} =
\begin{cases}
A(\bar{z})\left [r(\bar{z})\left(\frac{1}{3}\theta\,\delta_{ij}+\sigma_{ij}\right)n^in^j+\mathbf{d}\cdot \mathbf{n}\right]\left [ 1-\mathcal{H}(\bar{z})r(\bar{z})\right ] \, , & \quad  \text{inside Laniakea} \, , \\[1.5em]
A(\bar{z}) \left[ \mathbf{d}^{(\rm out)}\cdot \mathbf{n} \right] \left [ 1-\mathcal{H}(\bar{z})r(\bar{z})\right ] \, , & \quad \text{outside Laniakea} \, .
\end{cases}
\end{equation}
Notice that we model the external velocity field as a bulk motion $\mathbf{d}^{(\rm out)}$ , since the dipole is expected to provide the dominant contribution on scales larger than Laniakea. We assume a Gaussian likelihood
\be \label{eq:likelihood}
\log(\mathcal{L}) = - \frac{1}{2} \biggl[\Delta\mu^T \mathbf{C}^{-1} \Delta\mu + \log({\rm det}\ \mathbf{C})+k\log(2\pi)\biggr]\,,
\ee
where $k$ is the dimension of the vector of residuals $\Delta\mu$ defined by:

\begin{equation}
\label{e:cases}
\Delta \mu^i =
\begin{cases}
\mu^i + dM -\mu_\mathrm{ceph}^i,  \quad & i \in  \text{Cepheid hosts} \, ,\\[1em]
\mu^i +  dM -\mu_\mathrm{model}^i, \quad & \text{otherwise} \, ,
\end{cases}
\end{equation}

where $\mu^i_\mathrm{model}$ is given by~\eqref{eq:mu_model}, $\mu_\mathrm{ceph}^i$ is the distance modulus of the $i$-th Cepheid in the same galaxy as the $i$-th supernova, and $dM$ is a nuisance parameter sampled over and constrained by the supernovae in Cepheid-host galaxies. $\mathbf{C}$ is the full covariance matrix provided by the Pantheon+ dataset. When splitting the sample into supernovae located inside and outside the Laniakea region, the corresponding covariance matrices are obtained by consistently selecting the rows and columns of the full Pantheon+ covariance matrix associated with each subsample. This procedure preserves the statistical and systematic correlations between supernovae within each subset, ensuring that the propagated uncertainties are treated consistently throughout the analysis. 

We implemented our MCMC routine in {\tt JAX}~\footnote{Available at \url{https://github.com/jax-ml/jax}.}~\citep{Jax18repo, Jax18}, which provides differentiable and efficient {\tt{Python}} code optimized for GPUs. In particular, we have used the {\tt JAX-metal} version compatible with a Mac environment. The theoretical model described in Section~\ref{sec:theory} has been implemented in the code~\texttt{scoutpip}.~\footnote{Available at \url{https://github.com/fsorrenti/scoutpip}.}
As our MCMC sampler, we used a modified version of the {\tt JAX} implementation of {\tt affine},~\footnote{Available at \url{https://github.com/justinalsing/affine}.} which is a parallelized affine-invariant sampler based on {\tt emcee}~\cite{emcee}. We sampled the posterior distribution using 32 walkers and assuming wide flat priors. 

We used the integrated autocorrelation time $\tau$~\cite{autocorr} as a convergence diagnostic. $\tau$ can be interpreted as the number of steps the chain needs to forget where it started. In particular, convergence is achieved when the number of steps $N > 50\,\tau_{\rm max}$, with $\tau_{\rm max}$ being the maximum integrated autocorrelation time across all the parameters. We then removed $2 \, \lfloor \tau_{\rm max} \rfloor$ steps from the chain as burn-in \citep{burnin}. Finally, the chains are analysed using \texttt{chainconsumer v0.34} \citep{Hinton2016}.

\section{Results}\label{sec:results}

\begin{figure}
\centering
\includegraphics[scale=0.78]{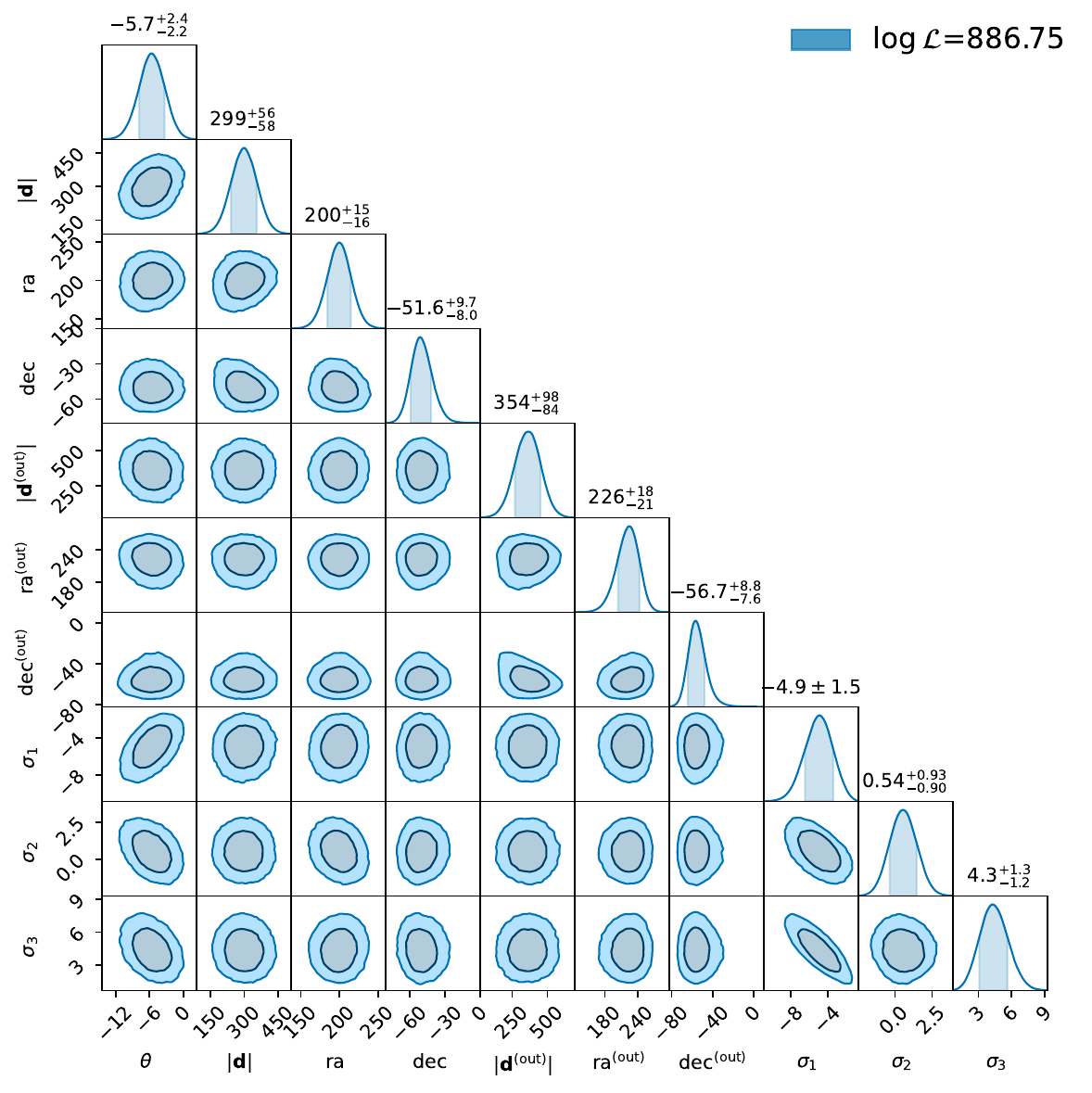}
\caption{
Triangle plots of the kinematic quantities constrained in this work using the Pantheon+ dataset.}
\label{fig:pantheon_kinematic_constraints}
\end{figure}

\begin{table*}[!htbp]
\centering
\renewcommand{\arraystretch}{1.35}

\begin{tabular}{|l|c|c|}
\hline\hline

Parameter
& Inside Laniakea
& Outside Laniakea \\[1.5mm]

\hline

Expansion $\theta$
& $-5.7^{+2.4}_{-2.2}$\,{\footnotesize km/s/Mpc}
& $-$ \\ [1.5mm]

\hline

Dipole amplitude $|\mathbf{d}|$
& $299^{+56}_{-58}$\,{\footnotesize km/s}
& $354^{+98}_{-84}$\,{\footnotesize km/s} \\

Dipole direction (ra)
& $200^{\circ}\,^{+15}_{-16}$
& $226^{\circ}\,^{+18}_{-21}$ \\

Dipole direction (dec)
& $-51.6^{\circ}\,^{+9.7}_{-8.0}$
& $-56.7^{\circ}\,^{+8.8}_{-7.6}$ \\ [1.5mm]

\hline

Shear $\sigma_1$
& $-4.9\pm1.5$\,{\footnotesize km/s/Mpc}
& $-$ \\

Shear $\sigma_2$
& $0.54^{+0.93}_{-0.90}$\,{\footnotesize km/s/Mpc}
& $-$ \\

Shear $\sigma_3$
& $4.3^{+1.3}_{-1.2}$\,{\footnotesize km/s/Mpc}
& $-$ \\ [1.5mm]

\hline
$e_1$ $(\mathrm{ra},\mathrm{dec})$
&
$(201^{\circ}\,^{+20}_{-15},\,53^{\circ}\,^{+14}_{-12})$
& $-$
\\

$e_2$ $(\mathrm{ra},\mathrm{dec})$
&
$(349^{\circ}\,^{+11}_{-33},\,29^{\circ}\,^{+14}_{-15})$
& $-$
\\

$e_3$ $(\mathrm{ra},\mathrm{dec})$
&
$(89^{\circ}\,^{+15}_{-13},\,17.8^{\circ}\,^{+9.6}_{-11.3})$
& $-$
\\[1.5mm]

\hline
$H_0$
& \multicolumn{2}{c|}{$73.6^{+1.1}_{-1.0}$\,{\footnotesize km/s/Mpc} } \\ [1.5mm]

\hline

$\Ommo$
& \multicolumn{2}{c|}{$0.332^{+0.019}_{-0.018}$} \\ [1.5mm]

\hline\hline

\end{tabular}
\caption{
Best-fit parameters for supernovae located inside and outside the Laniakea volume. Both cosmological parameters were constrained globally for the supernovae inside and outside Laniakea.
}
\label{tab:comparison}
\end{table*}

The best-fit parameters obtained from the joint likelihood analysis are
summarized in Tab.~\ref{tab:comparison}{\color{blue},} and the corresponding posterior distributions for the kinematic parameters are plotted in Fig.~\ref{fig:pantheon_kinematic_constraints}. The table reports the values of the
kinematic parameters inferred from the luminosity distance fit together
with the corresponding global cosmological parameters. Results are shown separately for supernovae located inside and outside the effective Laniakea volume.

The best-fit values for the cosmological parameters are in good agreement with the posterior of the Pantheon+ main analysis~\cite{Brout_2022} (i.e., $H_0=73.6$ and $\Ommo=0.334$). The fitted values of the expansion scalar are consistent with a negative effective expansion within the Laniakea region. The dipole amplitude, $|\mathbf{d}|$, shows a modest increase when considering supernovae located outside the Laniakea region, albeit with larger error bars. The dipole direction is broadly consistent between the two subsamples. It is interesting to note that the dipole is consistent within $2\sigma$ with the bulk motion found in CosmicFlows-4~\cite{Watkins_2023, Whitford_2023}, pointing close to the direction of the Shapley supercluster~\cite{Shapely:2006}. Finally, the shear eigenvalues inferred from the kinematic model display a clear anisotropic pattern inside the Laniakea volume.

In Sec.~\ref{subsec:kinematic_structure}, we showed that the Laniakea eigenvalues inferred by~\cite{Giani:2023aor} can be recovered from the Solar System position. The kinematic parameters inferred from the luminosity distance analysis (Tab.~\ref{tab:comparison})
are broadly consistent with the velocity field reconstruction of the
Laniakea region presented in \cite{Giani:2023aor}. In particular, the expansion scalar $\theta$ is found to be negative, indicating a convergent flow toward the dynamical center of the structure. The qualitative pattern of the shear eigenvalues is also preserved, yielding one positive value, one negative value, and one consistent with zero. However, the magnitudes of the
expansion and shear inferred from the supernova analysis appear
to be $\sim 2$ times larger than those obtained from the CosmicFlows–4
reconstruction.

The eigenvectors recovered from the supernova analysis are plotted in Fig.~\ref{fig:eigenvector_fit}. {Since eigenvectors are defined up to a sign, throughout the paper we report the direction pointing toward the Northern Hemisphere. Notably, they show a very good agreement with the directions of the ellipsoidal semi-axes modeled by \cite{Giani:2023aor}, which are summarized in Tab.~\ref{tab:laniakea_comparison}. This shows that the ellipsoidal modeling of Laniakea kinematics is well supported by supernova data independently and is confirmed by a reduced $\chi^2 = 0.89$ for $1701 - 15$ degrees of freedom (where $1701$ is the number of data points and $15$ is the number of fitted parameters).

\begin{figure}[t]
\centering
\begin{subfigure}{0.31\textwidth}
    \centering
    \includegraphics[width=\textwidth]{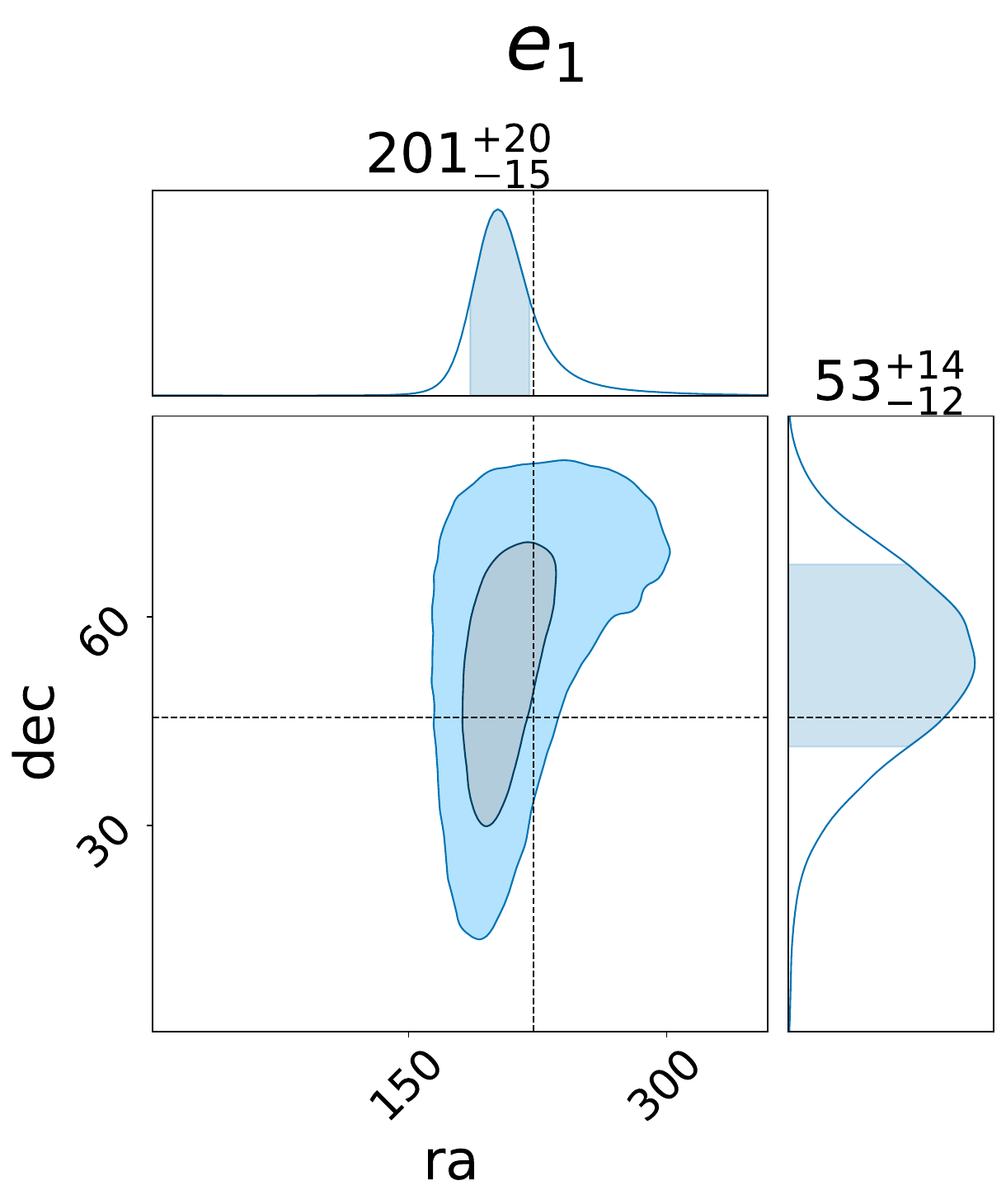}
    \label{fig:ev1}
\end{subfigure}
\begin{subfigure}{0.31\textwidth}
    \centering
\includegraphics[width=\textwidth]{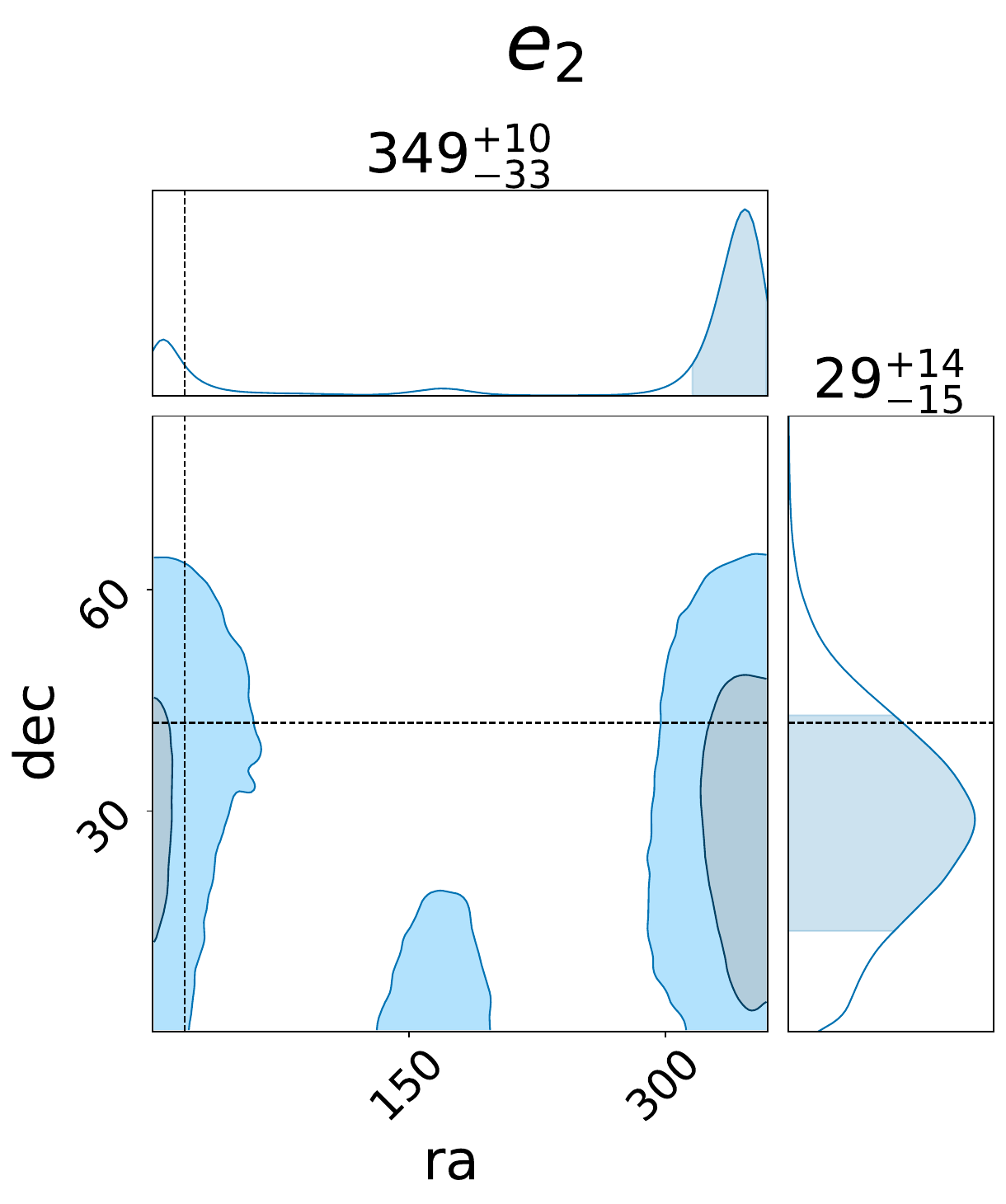}
    \label{fig:ev2}
\end{subfigure}
\begin{subfigure}{0.31\textwidth}
    \centering
    \includegraphics[width=\textwidth]{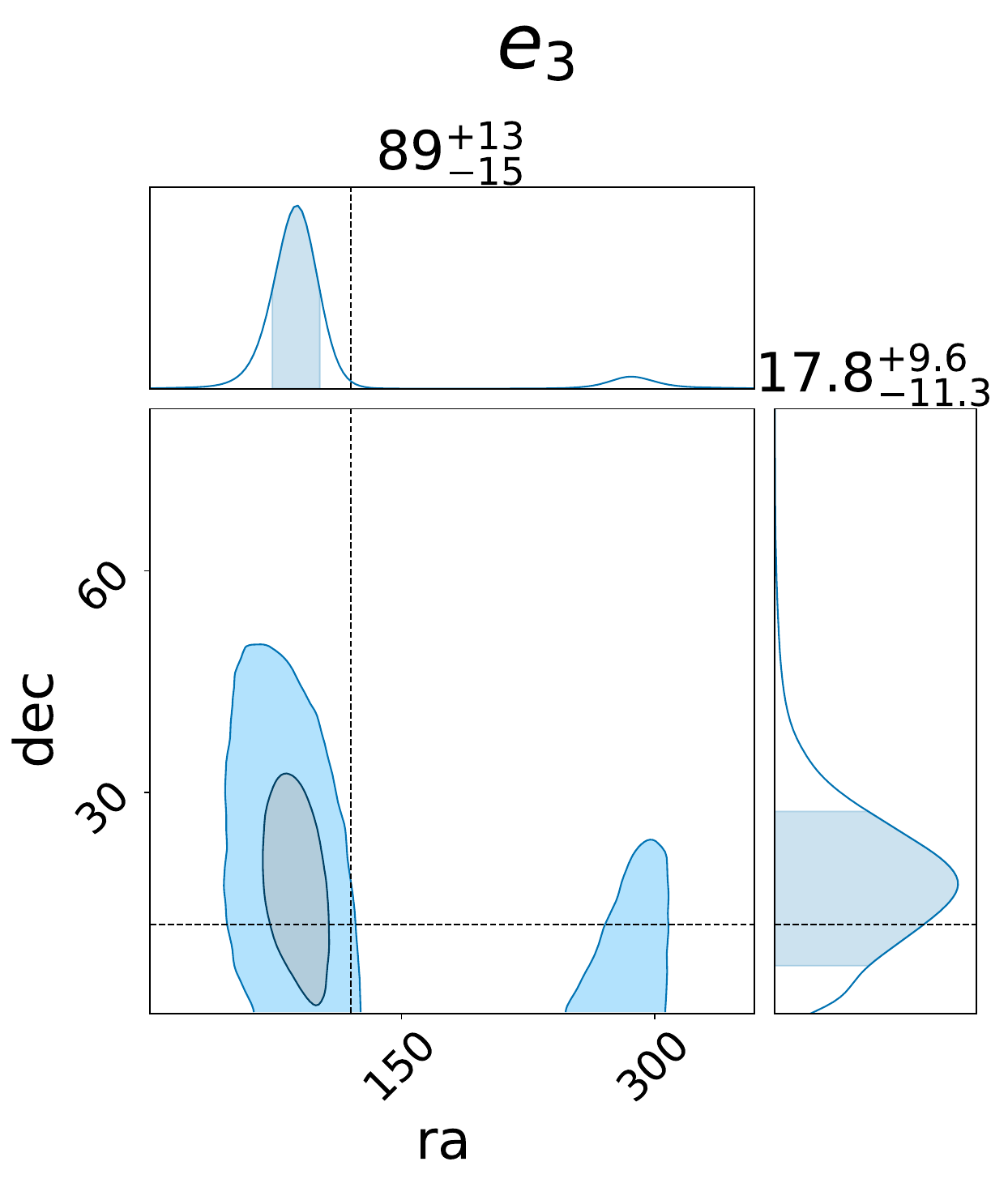}
    \label{fig:ev3}
\end{subfigure}
\caption{
Constraints on the eigenvector directions from our supernova analysis. The grey dashed lines show as a reference the directions of the ellipsoidal semi-axes assumed in~\cite{Giani:2023aor} and reported in Tab.~\ref{tab:laniakea_comparison}.}
\label{fig:eigenvector_fit}
\end{figure}

\begin{table*}[!htbp]
\centering
\renewcommand{\arraystretch}{1.35}
\begin{tabular}{|l|c|c|}
\hline\hline

Parameter
& This work
& Giani et al.~\cite{Giani:2023aor} \\[1.5mm]

\hline

Expansion $\theta$
&
$-5.7^{+2.4}_{-2.2}$\,{\footnotesize km/s/Mpc}
&
$-3.33 \pm 0.27$\,{\footnotesize km/s/Mpc}
\\[1.5mm]

\hline

Shear $\sigma_1$
&
$-4.9 \pm 1.5$\,{\footnotesize km/s/Mpc}
&
$-2.29 \pm 0.70$\,{\footnotesize km/s/Mpc}
\\

Shear $\sigma_2$
&
$0.54^{+0.93}_{-0.90}$\,{\footnotesize km/s/Mpc}
&
$0.46 \pm 0.55$\,{\footnotesize km/s/Mpc}
\\

Shear $\sigma_3$
&
$4.3^{+1.3}_{-1.2}$\,{\footnotesize km/s/Mpc}
&
$1.80 \pm 0.45$\,{\footnotesize km/s/Mpc}
\\[1.5mm]

\hline

$e_1$ $(\mathrm{ra},\mathrm{dec})$
&
$(201^\circ\,^{+20}_{-15},\,53^\circ\,^{+14}_{-12})$
&
$(222.66^\circ,\,45.52^\circ)$
\\

$e_2$ $(\mathrm{ra},\mathrm{dec})$
&
$(349^\circ\,^{+11}_{-33},\,29^\circ\,^{+14}_{-15})$
&
$(18.88^\circ,\,41.95^\circ)$
\\

$e_3$ $(\mathrm{ra},\mathrm{dec})$
&
$(89^{\circ}\,^{+15}_{-13},\,17.8^\circ\,^{+9.6}_{-11.3})$
&
$(120.02^\circ,\,12.13^\circ)$
\\[1.5mm]

\hline\hline

\end{tabular}
\caption{
Comparison between the kinematic parameters inferred in this work and the values reported in \citep{Giani:2023aor}. Our inferred eigenvector directions are compared with the directions of the ellipsoidal semi-axes assumed in~\cite{Giani:2023aor} and reported in Tab.~\ref{Tab:Lparameters}.
}
\label{tab:laniakea_comparison}
\end{table*}

\section{Mock validation}
\label{sec:mock_validation}

To assess the robustness of our method, we applied it to a mock SNe Ia dataset in which an ellipsoidal velocity field is injected as a redshift correction within a randomly oriented \emph{input} ellipsoidal region whose orientation is aligned with the velocity field itself. Outside this region, only a dipolar velocity field is injected. Using the same mock catalogue, we then repeat the analysis after selecting alternative geometries to determine whether a supernova is outside or inside the structure, in order to evaluate how the recovery of the kinematic parameters depends on the supernova selection. In particular, we consider geometries obtained by enlarging, rotating, and shifting the original input ellipsoid. More information about the mock analysis, including dataset characteristics and the ellipsoidal geometries{\color{blue},} can be found in Appendix~\ref{appB}. 

\begin{figure}
\centering
\includegraphics[width=1\textwidth]{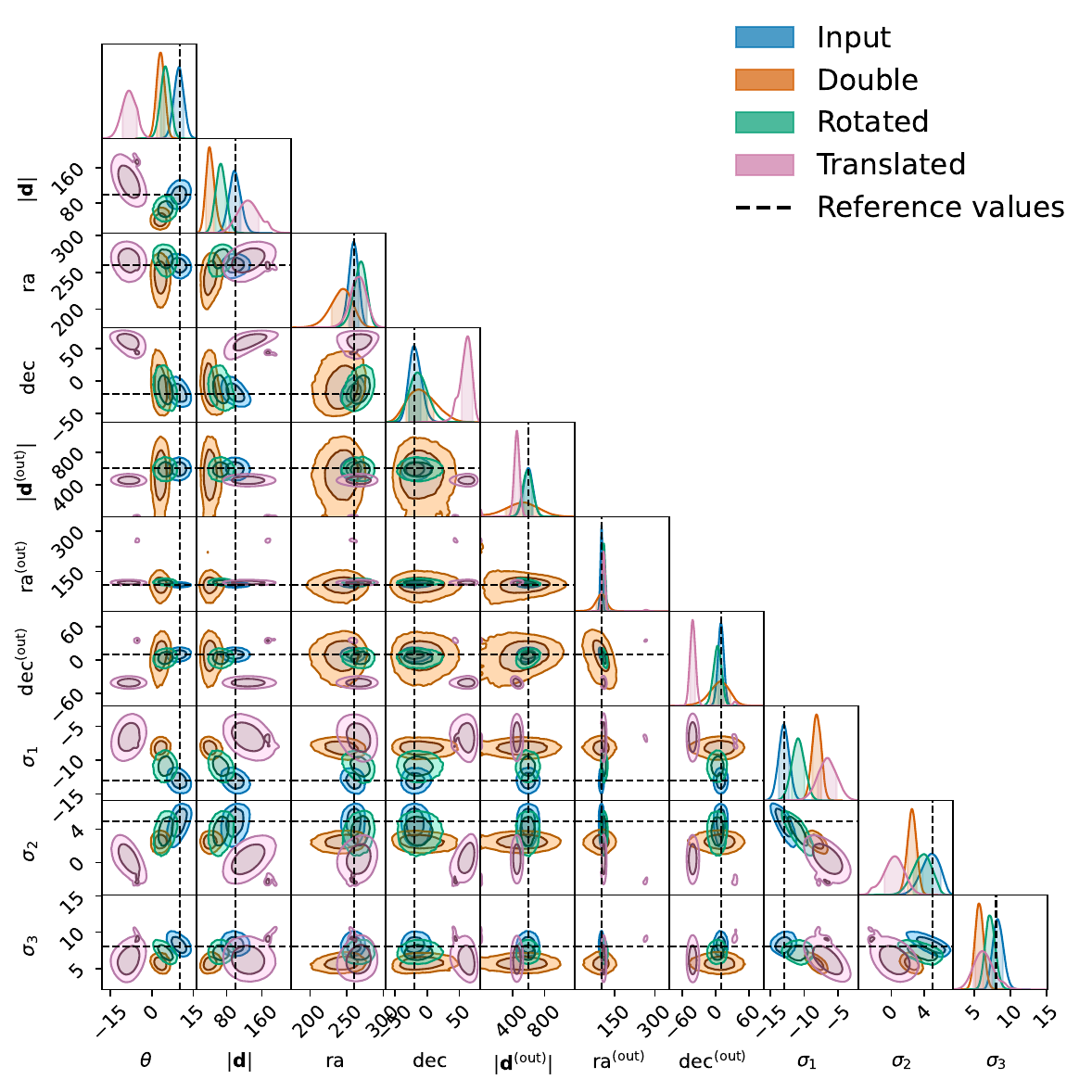}
\caption{
Constraints on the kinematic quantities for the mock dataset considering different geometries for the supernova selection.}
\label{fig:mock_kinematic_constraints}
\end{figure}

\begin{figure}[t]
\hspace{12pt} \includegraphics[scale=0.23]{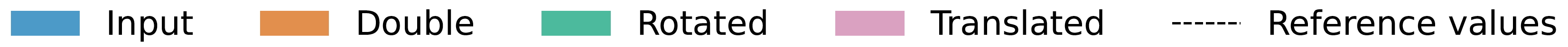}
\vspace{9pt}\\
\centering
\begin{subfigure}{0.31\textwidth}
    \centering
    \includegraphics[width=\textwidth]{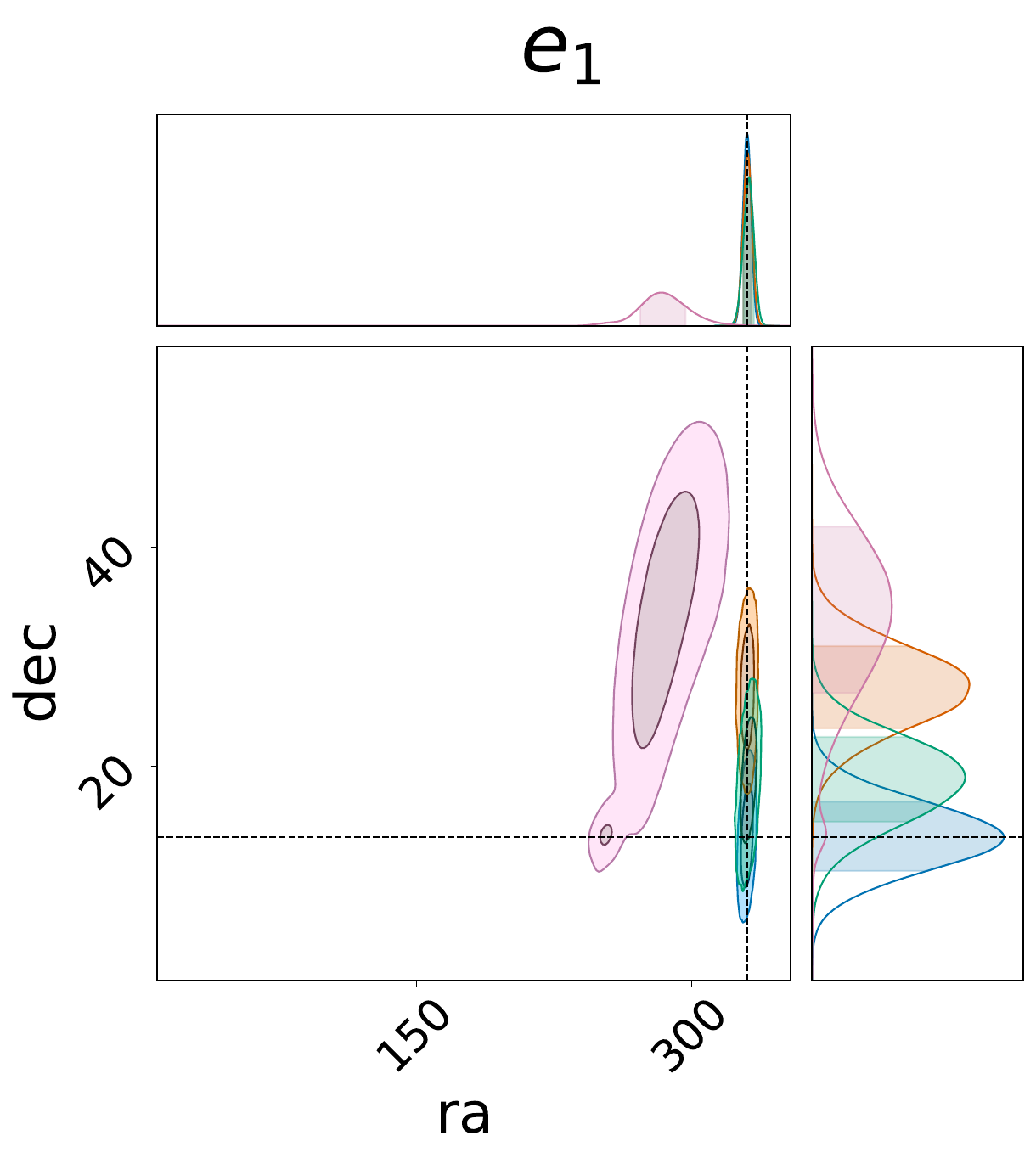}
    \label{fig:mock_ev1}
\end{subfigure}
\begin{subfigure}{0.31\textwidth}
    \centering
\includegraphics[width=\textwidth]{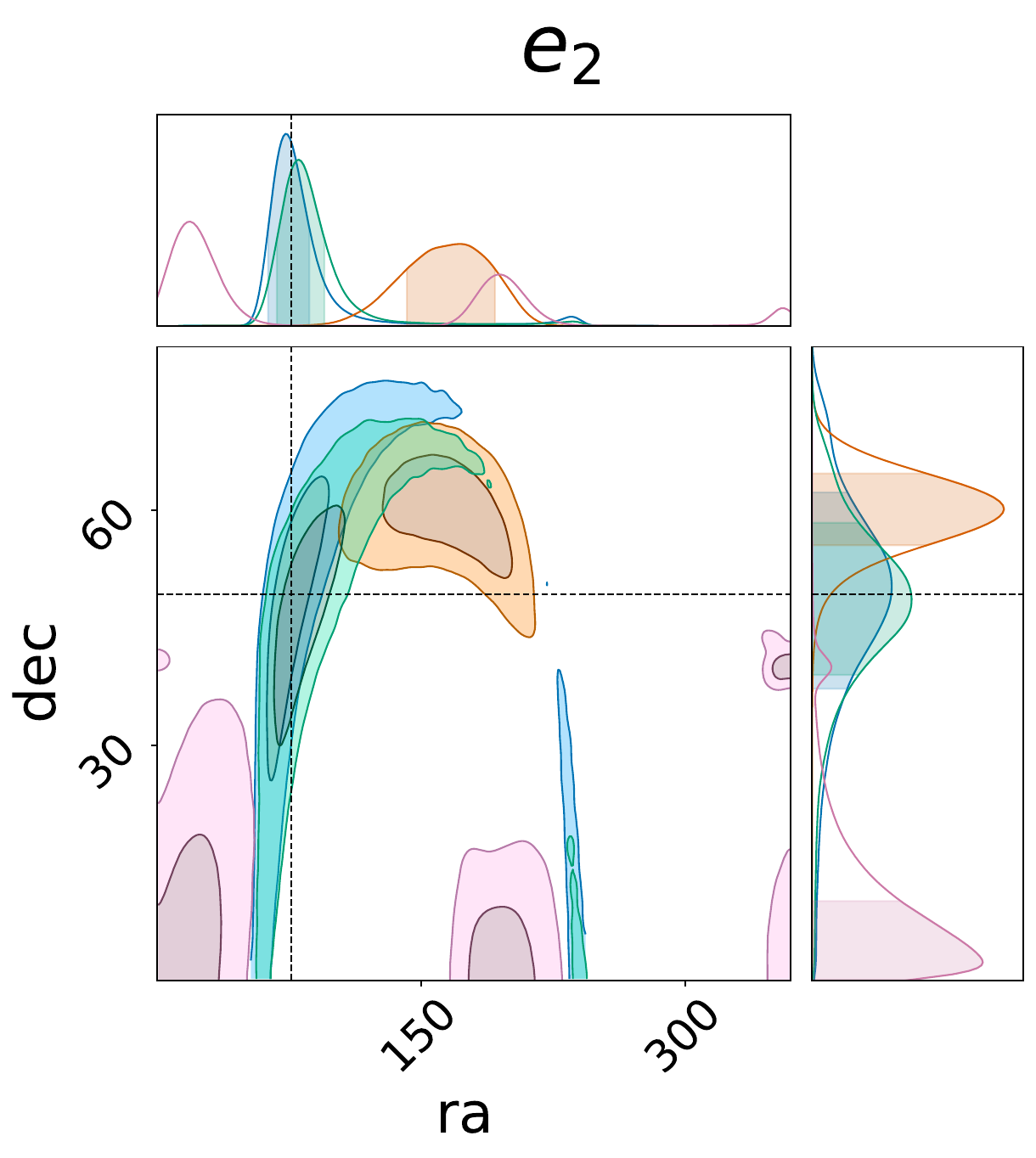}
    \label{fig:mock_ev2}
\end{subfigure}
\begin{subfigure}{0.31
\textwidth}
    \centering
    \includegraphics[width=\textwidth]{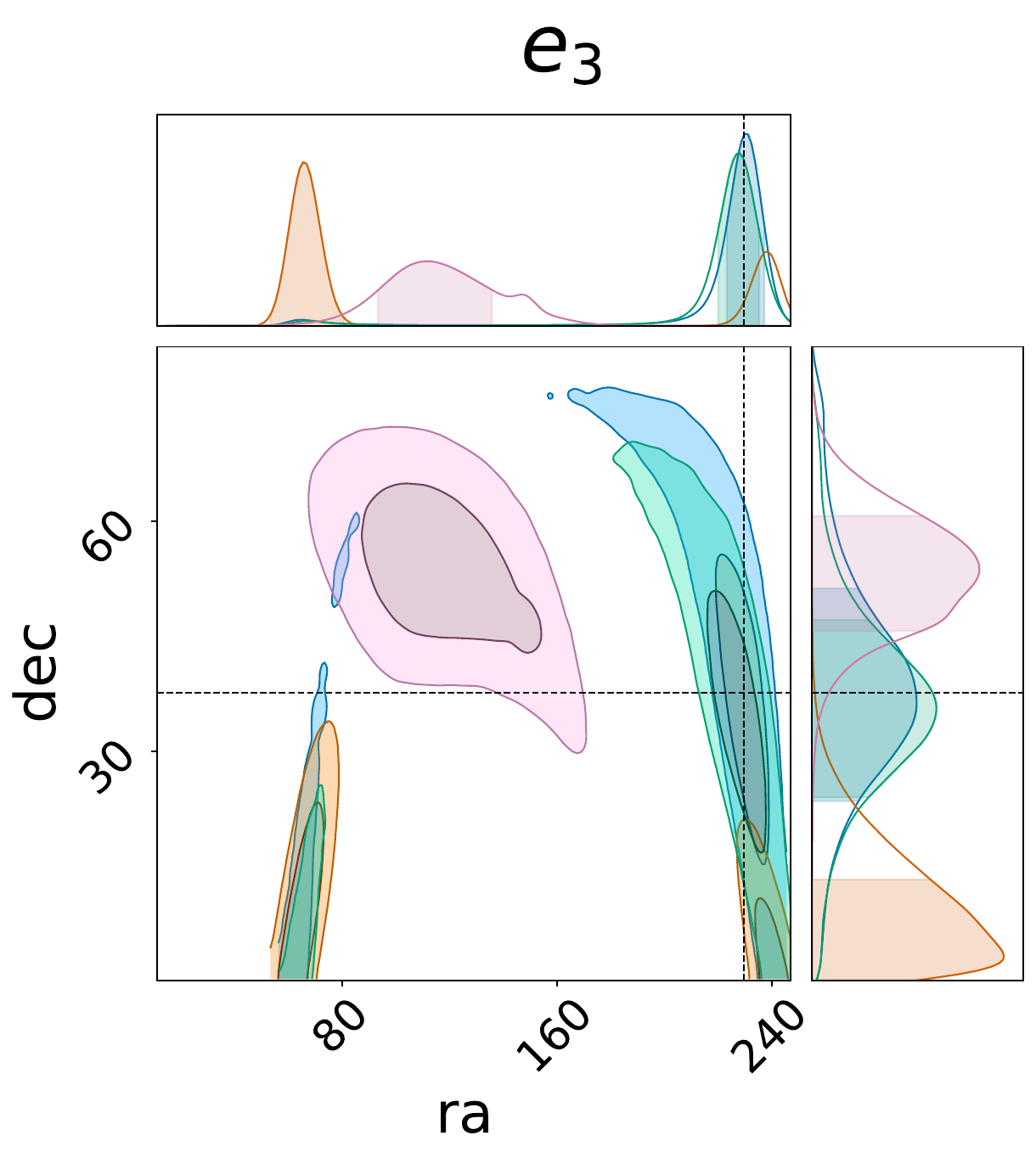}
    \label{fig:mock_ev3}
\end{subfigure}
\caption{
Constraints on the mock eigenvector directions for the different geometry selections.}
\label{fig:mock_ev}
\end{figure}

\begin{table*}[!htbp]
\centering
\renewcommand{\arraystretch}{1.35}
\begin{tabular}{|l|c|c|}
\hline\hline

Geometry selection & Number & Number of SNe Ia in common\\
 &of SNe Ia&with the input geometry\\
 [1.5mm]
\hline

Input & 302 & 302 \\ 

Double semiaxes & 577 & 302 \\ 

Rotated & 301 & 242 \\

Translated & 166 & 61 \\
[1.5mm]

\hline\hline

\end{tabular}

\caption{
The second column shows the number of mock SNe Ia selected by each geometry. The third column reports the number of supernovae in common with the input geometry; from this, we can easily infer that in the case of the enlarged and translated geometries, the selected samples include a substantial number of supernovae outside the region where the ellipsoidal velocity field was injected. 
}\label{tab:numbers_mock}

\end{table*}

The best-fit kinematic parameters obtained for the different mock geometries are shown in Fig.~\ref{fig:mock_kinematic_constraints}. The recovered eigenvector directions are shown in Fig.~\ref{fig:mock_ev}, together with the input directions used as reference. From these figures, we see that when the supernova selection matches the input ellipsoidal region, the method accurately recovers the injected kinematic parameters, including the expansion, shear eigenvalues, dipole amplitude, and eigenvector directions. This confirms that the luminosity distance multipoles contain enough information to reconstruct the imposed ellipsoidal velocity field when the selected volume traces the region where the field is actually injected.

From Tab.~\ref{tab:numbers_mock}, we can see that in the case of the enlarged and translated geometries, the selected samples include a substantial number of supernovae outside the region where the ellipsoidal velocity field was injected. As a result, the signal is diluted, producing significant biases in the inferred parameters, as we can see in Fig.~\ref{fig:mock_kinematic_constraints}. 
This shows that an incorrect size of the volume selection can strongly suppress the inferred anisotropic signal. Moreover, this dilution can also bias the cosmological parameter inference. In particular, as shown in Fig.~\ref{fig:mock_theta_cosmo}, $\theta$ exhibits a negative degeneracy with $H_0$ and a positive degeneracy with $\Ommo$; therefore, a bias in $\theta$ translates into a bias in the cosmological parameters.

\begin{figure}
\centering
\includegraphics[scale=0.4]{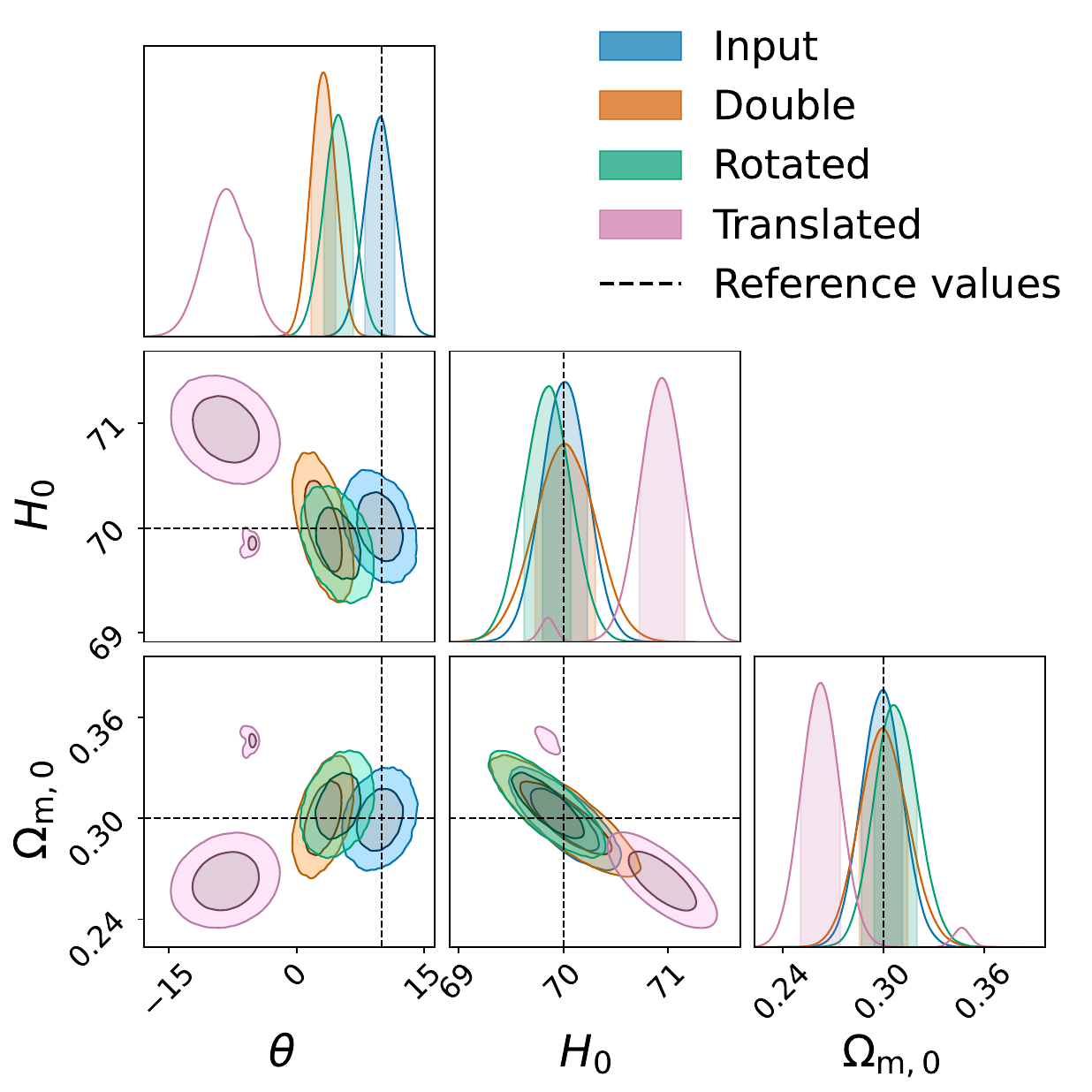}
\caption{
Constraints on the expansion $\theta$ and the cosmological parameters $H_0$ and $\Ommo$ for the mock analysis. We can see that $\theta$ has a negative degeneracy with $H_0$ and a positive degeneracy with $\Ommo$.}
\label{fig:mock_theta_cosmo}
\end{figure}

On the other hand, the rotated geometry produces milder biases in the recovered amplitudes. In particular, the expansion and shear eigenvalues are shifted with respect to the input values, indicating that the inferred magnitudes are sensitive to the spatial overlap between the selected geometry and the true region supporting the velocity field. Nevertheless, the recovered eigenvector directions remain remarkably stable in the rotated selection. This stability suggests that the method is especially robust in recovering the principal directions of the underlying ellipsoidal flow when a significant number of external supernovae is not introduced, as was the case for the selected rotation.

\section{Conclusions}\label{sec:conclusions}

In this work, we developed a kinematic framework that relates the monopole and quadrupole components of the luminosity distance to the expansion and shear of a local peculiar velocity field. Applying this formalism to SNe Ia located within the effective ellipsoidal volume of Laniakea, we recover a negative expansion scalar, an anisotropic shear pattern, and a set of eigenvector directions consistent with the asymptotic projections of the principal axes of the supercluster. These results are qualitatively consistent with the reconstruction of the Laniakea velocity field obtained from CosmicFlows-4 \cite{Giani:2023aor}, including the sign of the expansion and the relative structure of the shear eigenvalues. We also inferred a dipole consistent with the bulk motion found independently in CosmicFlows-4~\cite{Watkins_2023, Whitford_2023}, pointing close to the direction of the Shapley supercluster~\cite{Shapely:2006}. The agreement between the supernova reconstruction and the CosmicFlows-4 analysis suggests that luminosity distance anisotropies contain direct information about the underlying kinematics of nearby large-scale structures.

Using mock supernova catalogues in which an ellipsoidal peculiar velocity field was injected within a randomly oriented ellipsoidal volume, we demonstrated that the underlying kinematics can be accurately recovered when the correct geometry is used to select supernovae inside the kinematic region. Conversely, adopting an incorrect geometry introduces systematic biases in both kinematic and cosmological parameters. Interestingly, despite these biases, the reconstructed eigenvector directions remain stable provided the selection geometry does not include a large number of supernovae outside the true input region, indicating that the orientation of the underlying velocity field is more robustly constrained than the amplitudes of its kinematic components.

It is important to note that, as a first approximation, the velocity field was modelled as an ellipsoidal flow only within the Laniakea volume, while the external region was described by a purely dipolar component. Recent reconstructions of the local velocity field suggest that the dynamics outside the supercluster generally contain additional contributions beyond a simple bulk flow description \cite{Hoffman2023,Watkins2026}. Future extensions of this framework could therefore incorporate a more realistic external velocity field and replace the sharp interior/exterior classification with a continuous transition function across the effective boundary of the structure, reflecting the fact that the underlying velocity field is expected to vary smoothly in space. 

To summarize, at present, the large-scale structure of the local Universe is primarily mapped through galaxy density reconstructions and peculiar velocity catalogues based on distance indicators. In this work, we have shown that anisotropies in the luminosity distance provide a complementary and independent route to describe the kinematics of nearby structures directly from supernova observations. This methodology is particularly relevant in the context of forthcoming surveys such as ZTF and LSST. In particular, the number of SNe Ia expected from LSST will exceed current samples by orders of magnitude, substantially improving both sky coverage and statistical precision \cite{Ivezi__2019,thelsstdarkenergysciencecollaboration2021lsstdarkenergyscience}. Under these conditions, in the near future it will become possible to reconstruct the expansion and shear of multiple nearby superstructures using only SNe Ia.

%%%%%%%%%%%%%%%%%%%%%%%%%%%%%%%%%%%%%%%%%%%%%

\acknowledgments
FS acknowledges support from the Swiss National Science Foundation (SNSF) through the Postdoc.Mobility fellowship (grant no. P500-2\_235508). EP acknowledges support from the POSTDOC\_DICYT project 042531CM\_Postdoc, Vicerrectoría de Investigación, Innovación y Creación, Universidad de Santiago de Chile (USACH). LG acknowledges support from the Australian Government through the Australian Research Council Centre of Excellence for Gravitational Wave Discovery (OzGrav).
%%%%%%%%%%%%%%%%%%%%%%%%%%%%%%%%%%%%%%%%%%%%%
%%%%%%%%%%%%%%%%% APPENDICES %%%%%%%%%%%%%%%%%%%%%

\appendix

\section{Ellipsoidal expansion of the peculiar velocity field} 
\label{app0}

The velocity of a source with respect to the Laniakea frame is related to its velocity measured in the Solar System frame as
\begin{eqnarray}
    \mathbf{v}_{\rm src/L}=\mathbf{R}\cdot \mathbf{v}_{\rm src/\odot}+\mathbf{v}_{\rm \odot/L}\,,
\end{eqnarray}
Inverting this relation yields:
\begin{eqnarray}
    \mathbf{v}_{\rm src/\odot}=\mathbf{R}^{-1}\cdot(\mathbf{v}_{\rm src/L}-\mathbf{v}_{\rm \odot/L})\,.
\end{eqnarray}
Boosting this velocity to the CMB frame, we obtain the peculiar velocity of the source with respect to the Hubble flow
\begin{eqnarray}
    \mathbf{v}_{\rm src/CMB}&=&\mathbf{R}^{-1}\cdot(\mathbf{v}_{\rm src/L}-\mathbf{v}_{\rm \odot/L})+\mathbf{v}_{\rm \odot/CMB}\,,
    \label{eqn:velocitysourceCMB}
\end{eqnarray}
where $\mathbf{v}_{\rm \odot/CMB}$ is the velocity of the Solar System with respect to the CMB. Describing the velocity field in the Laniakea frame using the diagonal tensor $\mathbf{\Delta H}$, we have that
\begin{eqnarray}
    \mathbf{v}_{\rm src/L}=\mathbf{\Delta H} \cdot \mathbf{x}_{\rm src/L}\,.
\end{eqnarray}
The peculiar velocity field as inferred from the Solar System reads
\begin{eqnarray}\label{Aeq:source_intermediate}
    \mathbf{v}_{\rm src/\odot}=\mathbf{R}^{-1}\cdot (\mathbf{\Delta H} \cdot \mathbf{x}_{\rm src/L} - \mathbf{v}_{\rm \odot/L})\,.
\end{eqnarray}
Substituting Eq.~\eqref{positions} in Eq.~\eqref{Aeq:source_intermediate}, we obtain:
\begin{eqnarray}
    \mathbf{v}_{\rm src/\odot}
    &=&\mathbf{R}^{-1}\cdot \left[\mathbf{\Delta H} \cdot (\mathbf{R}\cdot \mathbf{x}_{\rm src/\odot}+\mathbf{p}) - \mathbf{v}_{\rm \odot/L}\right] \\
    &=& \mathbf{E}\cdot \mathbf{x}_{\rm src/\odot}+\mathbf{K}\,, \label{Aeq:decomposition}
\end{eqnarray}
where we have defined
\begin{eqnarray}
    \mathbf{E}&=&\mathbf{R}^{-1}\cdot \mathbf{\Delta H} \cdot \mathbf{R},\label{Aeq:similarity}\\
    \mathbf{K}&=&\mathbf{R}^{-1}\cdot (\mathbf{\Delta H} \cdot \mathbf{p}- \mathbf{v}_{\rm \odot/L})\,.
\end{eqnarray}
This expression provides a model for the peculiar velocity field observed from the Solar System when the local dynamics of Laniakea is described by an ellipsoidal flow. From the Solar System perspective, the velocity field naturally decomposes into an ellipsoidal component, encoded in $\mathbf{E}$, and an additional dipolar contribution described by the vector $\mathbf{K}$. Using Eq.~\eqref{Aeq:decomposition}, the peculiar velocity of a source with respect to the CMB can be written in terms of the ellipsoidal field $\mathbf{E}$ as
\begin{eqnarray}
    \mathbf{v}_{\rm src/CMB}=\mathbf{v}_{\rm src/\odot}+\mathbf{v}_{\rm \odot/CMB}=\mathbf{E}\cdot\mathbf{x}_{\rm src/\odot}+\mathbf{d}\,,
    \label{Apeculiarvelocityfield_nored}
\end{eqnarray}
where $\mathbf{d}=\mathbf{K}+\mathbf{v}_{\rm \odot/CMB}$ is the dipolar component of the velocity field measured from the CMB rest frame.

Until now, we have neglected the redshift dependence of $\mathbf{v}_{\rm src/\odot}$. Taking the peculiar velocity redshift dependence from Eq.~\eqref{theory:v_p}, we have that Eq.~\eqref{Apeculiarvelocityfield_nored} becomes:
\begin{eqnarray}
    \mathbf{v}_{\rm src/CMB}=A(\bar{z})\,\left[\mathbf{E}\cdot\mathbf{x}_{\rm src/\odot}+\mathbf{d}\right]%\left[ 1-\mathcal{H}(\bar{z})r(\bar{z})\right]\,,
    \label{Apeculiarvelocityfield}
\end{eqnarray}
The prefactor $A(\bar{z})$ accounts for the time evolution of peculiar velocities. It is defined by
\begin{eqnarray}
    \label{theory:prefactor}
A(\bar{z})=\frac{D_1(\bar{z}) f(\bar{z})H(\bar{z})}{(1+\bar{z}) D_1(0) f(0) H_0} \, ,
\end{eqnarray} 
and it is typically $\sim 1$ at low redshift. Assuming a flat $\Lambda$CDM Universe, we can use the analytical expressions~\cite{Euclid:2023qyw}
\be \label{theory:fz1}
    D_1(\bar{z})=\frac{1}{5 \, (1+\bar{z}) \, \Ommo}\left[ _2F_1 \left ( \frac{1}{3},1;\frac{11}{6};1-\frac{1}{\Om_\text{m}(\bar{z})} \right) \right],
\ee
for the linear growth function and
\be \label{theory:fz}
    f(\bar{z})=\frac{1}{2}\Om_\text{m}(\bar{z}) \left[ \frac{5}{_2F_1 \left ( \frac{1}{3},1;\frac{11}{6};1-\frac{1}{\Om_\text{m}(\bar{z})} \right) } - 3 \right],
\ee
for the growth rate, where
\be
\Omm(\bar{z})=\frac{\Ommo \, (1+\bar{z})^{3}}{\Ommo \, (1+\bar{z})^{3}+(1-\Ommo)}\, ,
\ee
and $_2F_1(a,b;c;d)$ is the confluent hypergeometric function~\cite{Abra}.

\begin{figure*}[t]
\centering
\includegraphics[scale=0.65]{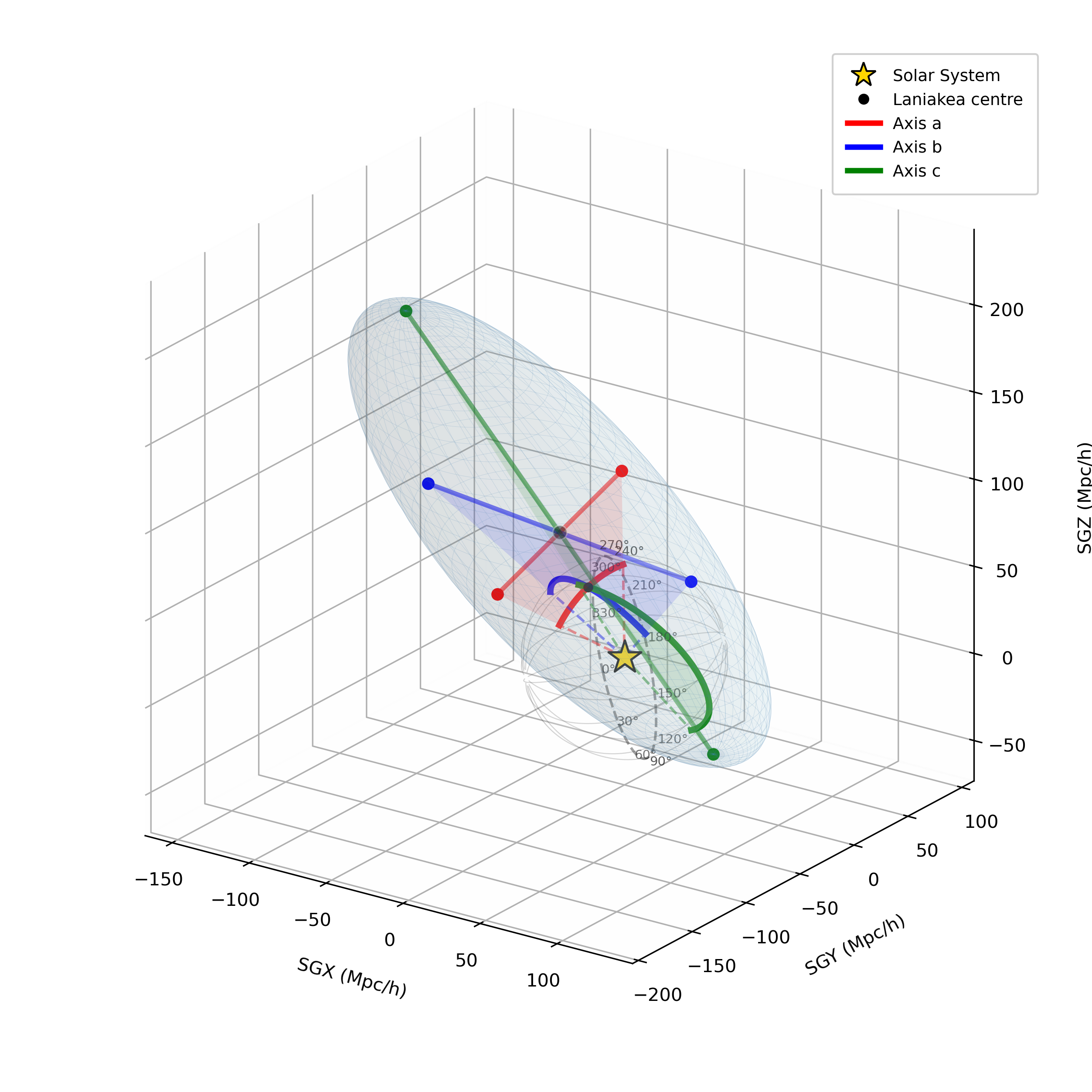}
\caption{Three-dimensional representation of the principal axes of the Laniakea ellipsoid and their projection onto the celestial sphere as observed from the Solar System in equatorial coordinates. 
For each principal axis, the projection plane defined by the Solar System and the physical axis of the ellipsoid generates a finite angular arc on the celestial sphere, shown by the coloured curves. }
\label{fig:laniakea_celestial_projection}
\end{figure*}

\begin{figure}
\centering
\includegraphics[scale=0.65]{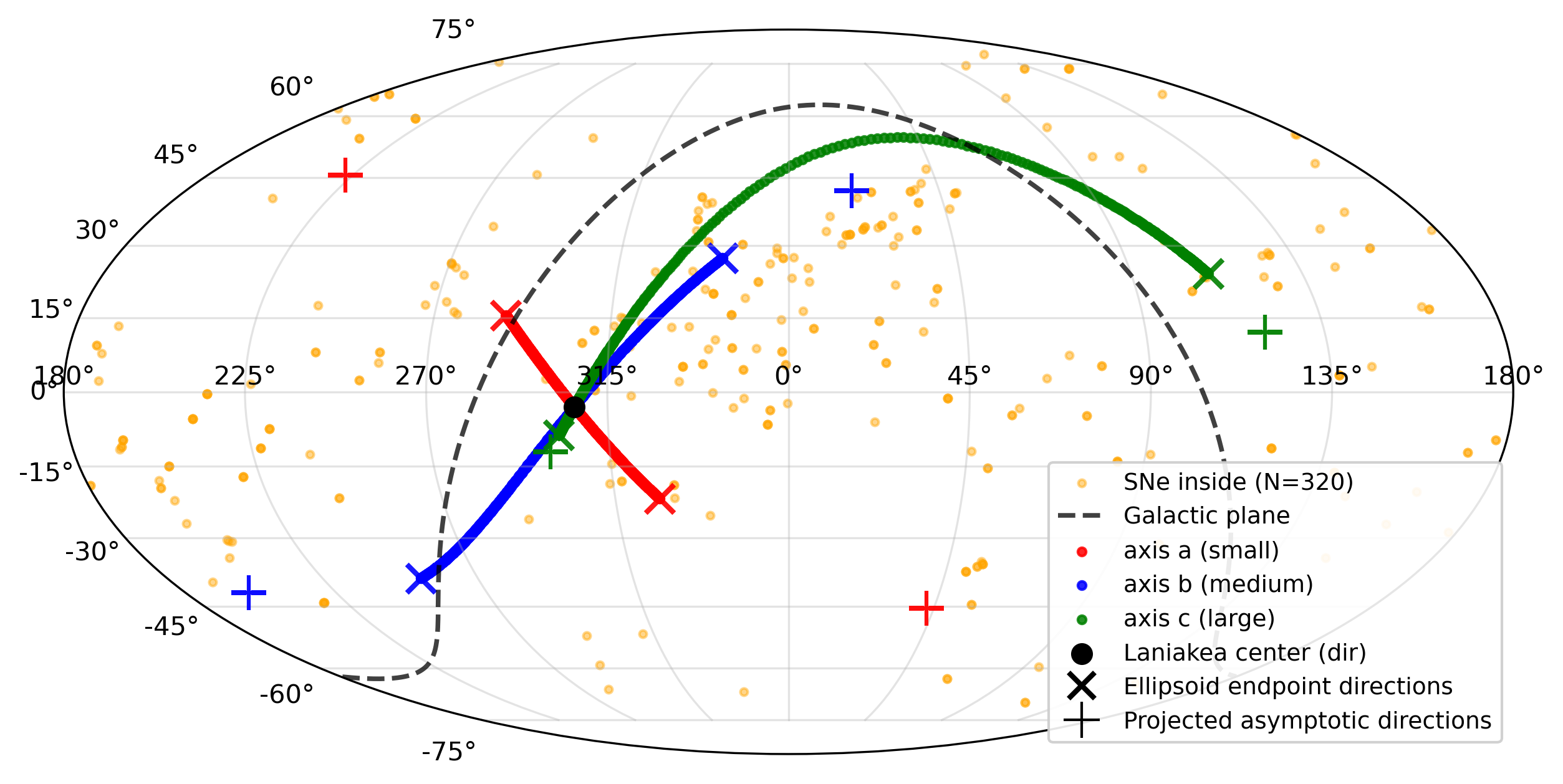}
\caption{
Projection of the principal axes of the Laniakea ellipsoidal velocity
field on the celestial sphere together with the positions of the
Pantheon+ supernovae inside the ellipsoid. Because the observer is not located
at the dynamical centre of the structure, the principal axes project
onto extended curves and not fixed directions on the sky.}
\label{fig:laniakea_axes_projection}
\end{figure}

\section{Laniakea eigenvectors and ellipsoidal axis} 
\label{appA}

The eigenvectors $\hat{\mathbf e}_i
$  measured in the Solar System do not correspond to the finite endpoints of the Laniakea ellipsoid (see Figs. \ref{fig:laniakea_celestial_projection} - \ref{fig:laniakea_axes_projection}). Indeed, from the Solar System position, the finite endpoints of a given axis are simply the sum between the vector position $\mathbf x_{\rm L}$ representing the Laniakea center and the vector that goes from the center to the finite endpoints of the ellipsoid, i.e.{\color{blue}, }
\begin{equation}
\mathbf X_{\pm}
=
\mathbf x_{\rm L}
\pm
a_i \hat{\mathbf e}_i\,,
\end{equation}
where $a_i$ denotes the corresponding semi-axis length. Because the observer is displaced from the centre of the structure, the angular positions of these endpoints depend on both $\mathbf x_{\rm L}$ and $a_i$.  On the other hand, the eigenvectors, which represent directions rather than points, can be obtained by considering the asymptotic limit
\begin{equation}
\mathbf X_{\pm}(r)
=
\mathbf x_{\rm L}
\pm
r\hat{\mathbf e}_i\, .
\end{equation}
with $r\rightarrow\infty$. The corresponding line-of-sight direction measured by the observer is
\begin{equation}
\lim_{r\rightarrow\infty}
\frac{\mathbf X_{\pm}(r)}
{\left|\mathbf X_{\pm}(r)\right|}
= \lim_{r\rightarrow\infty}
\frac{r
\left (\frac{\mathbf x_{\rm L}}{r}
\pm
\hat{\mathbf e}_i \right )}
{r\sqrt{\frac{\left|\mathbf x_{\rm L}\right|^2}{r^2}
\pm
\frac{2\mathbf e_i \mathbf x_{\rm L}}{r}+\left|\hat{\mathbf e}_i\right|^2}}=
\pm\hat{\mathbf e}_i \,.
\label{eq:eigenvectorequivalence}
\end{equation}
Therefore, the eigenvectors of the shear tensor measured in the Solar System frame have to be interpreted as the asymptotic directions obtained by extending the principal axes of the ellipsoid to infinity.

\section{Mock analysis details}
\label{appB}

\begin{figure}[t]
\centering

\begin{subfigure}{0.47\textwidth}
    \centering
    \includegraphics[width=\textwidth]{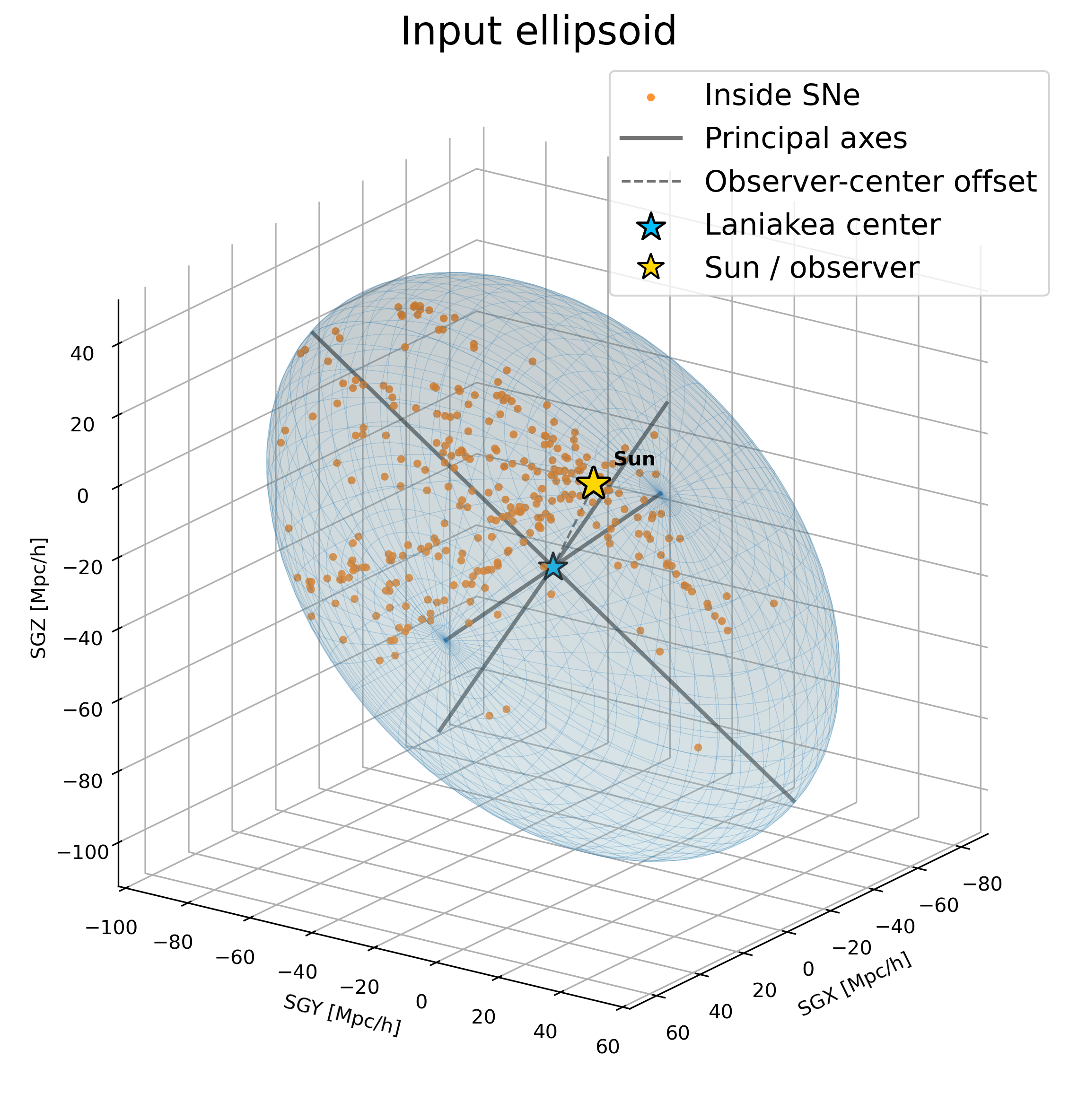}
    \label{fig:input_ellipsoid}
\end{subfigure}
\hfill
\begin{subfigure}{0.47\textwidth}
    \centering
    \includegraphics[width=\textwidth]{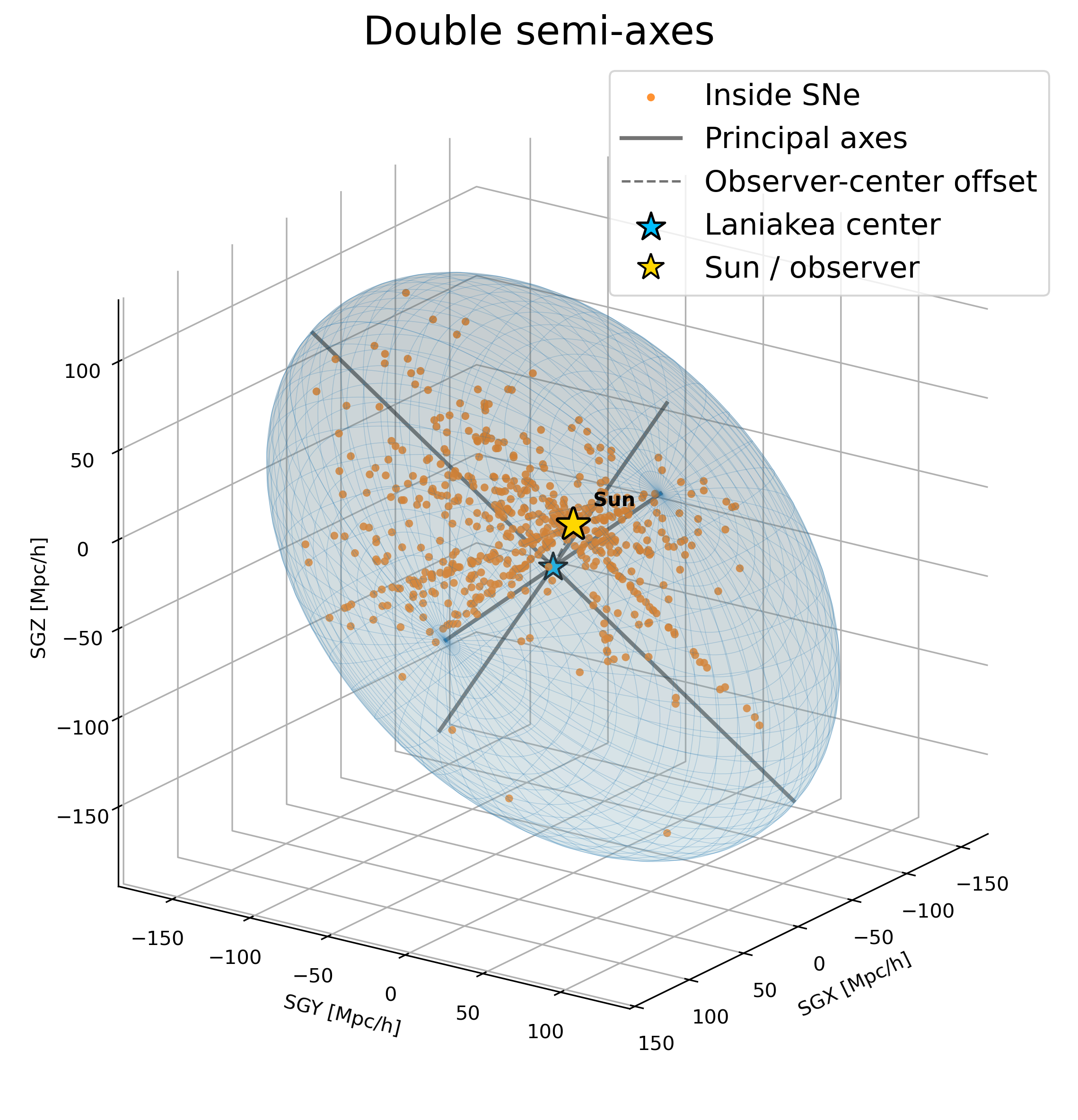}    \label{fig:double_ellipsoid}
\end{subfigure}

\begin{subfigure}{0.47\textwidth}
    \centering
    \includegraphics[width=\textwidth]{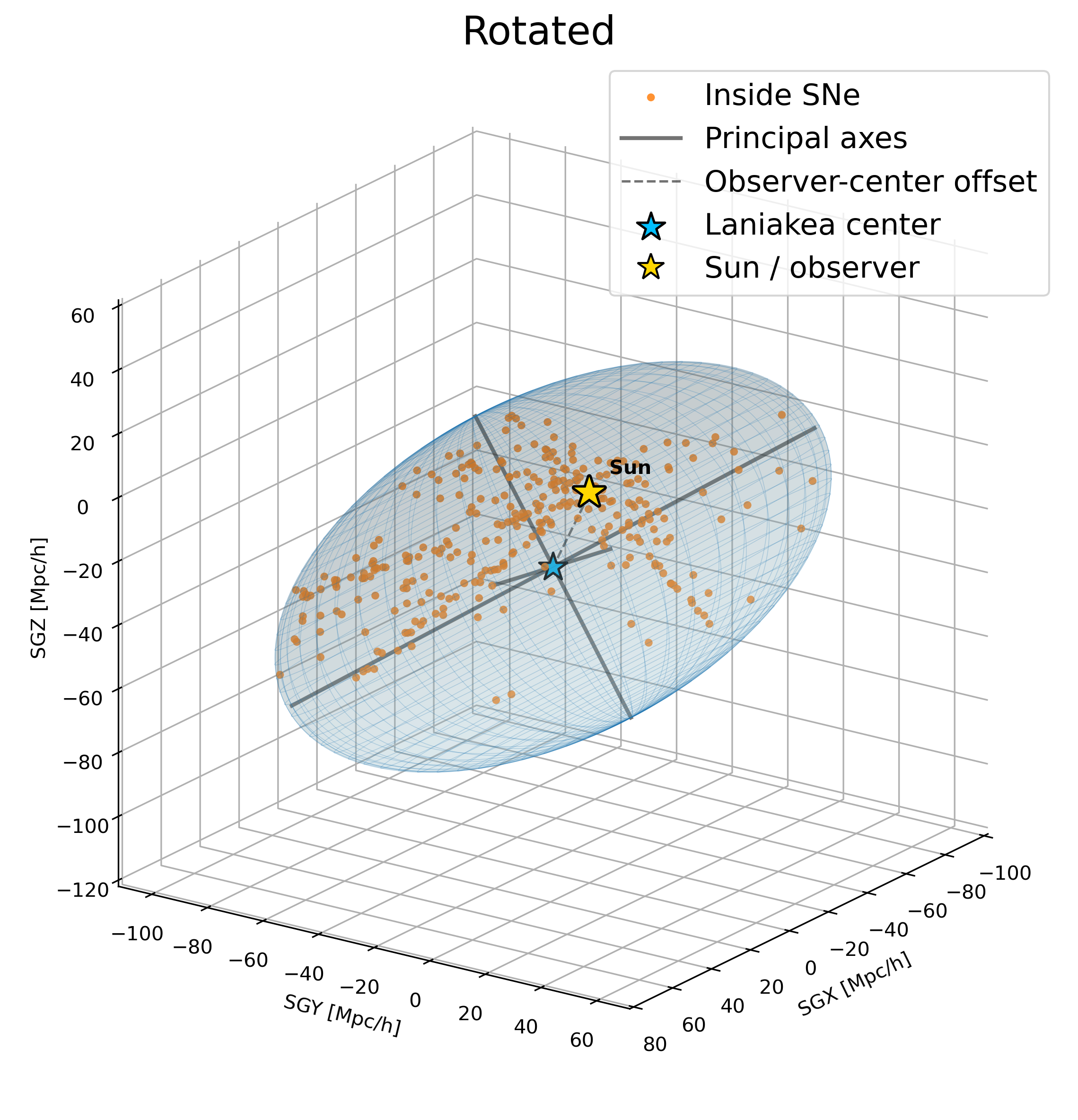}
    \label{fig:rotated_ellipsoid}
\end{subfigure}
\hfill
\begin{subfigure}{0.47\textwidth}
    \centering
    \includegraphics[width=\textwidth]{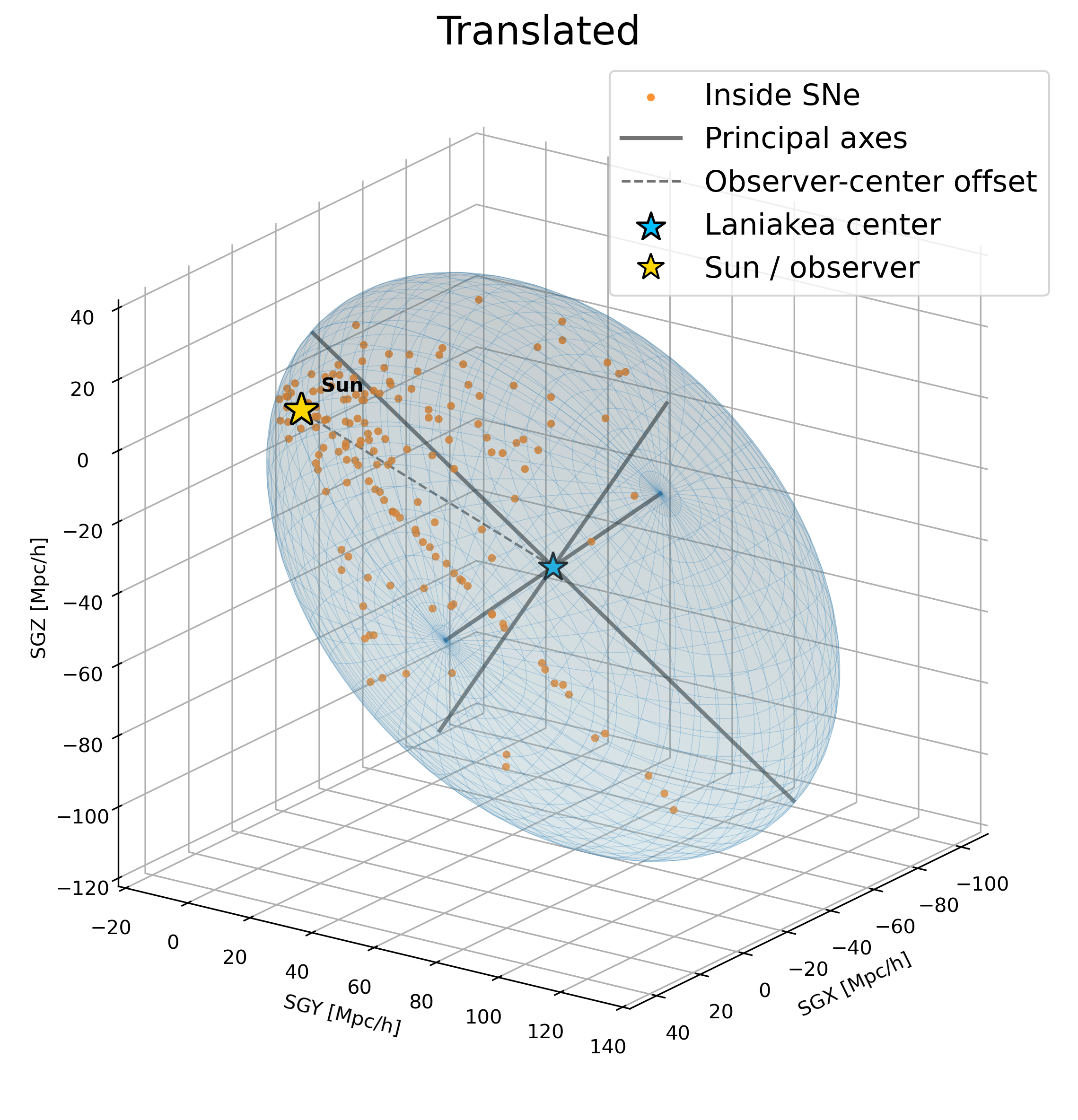}
    \label{fig:traslated_ellipsoid}
\end{subfigure}

\caption{
Spatial distribution of mock SNe Ia within the different mock geometries. The input ellipsoid corresponds to the geometry in which an ellipsoidal velocity field, with the parameters listed in Tab.~\ref{tab:input_mock}, is injected. We also analyse enlarged, rotated, and shifted versions of this geometry to assess the robustness of our method
}
\label{fig:mock_sn_distribution}
\end{figure}

\begin{figure}[H]
	\centering
	\includegraphics [scale=0.7]{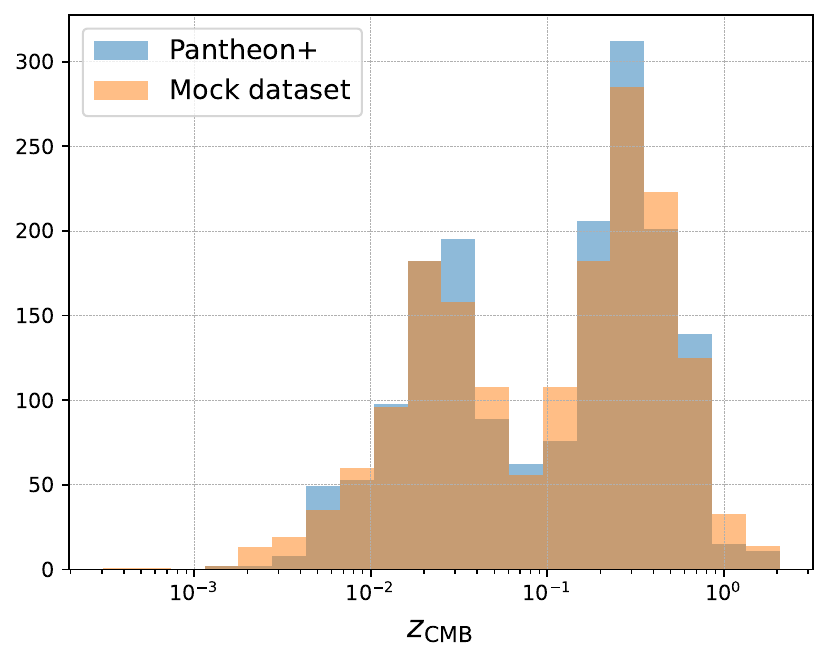}
\caption{Distributions of the CMB corrected redshifts $\bar{z}$ for the Pantheon+ dataset and the input mock dataset.\label{f:pantheon_plus_mocks}} 
\vspace{0.17cm}
\end{figure}

To create the mock dataset, we firstly substituted the CMB corrected redshifts $\bar{z}$ using a random redshift distribution with a similar shape, as we can see in Fig.~\ref{f:pantheon_plus_mocks}. Keeping the same supernovae angular distribution (ra, dec), we then computed the distance moduli according to the theoretical model of Sec.~\ref{sec:theory} using the input parameters of Tab.~\ref{tab:input_mock}. Finally, we applied our MCMC routine. For simplicity, we used a diagonal covariance whose elements are given by the diagonal of the original Pantheon+ covariance matrix. We did not consider supernovae in Cepheid-host galaxies, fixing $dM=0$.

The geometries used to inject a fiducial velocity field in supernovae mock data are shown in Fig.~\ref{fig:mock_sn_distribution}. In particular, we used an input ellipsoid with semi-axis of lenghts $90,\> 70$ and $50\> \mathrm{Mpc}/h$, while its directions are described by the eigenvectors of Tab.~\ref{tab:input_mock}. The enlarged geometry was made simply doubling each of its semi-axis. For the rotated geometry, we use a composite rotation on each supergalactic axis of $110^\circ$ in SGX, $70^\circ$ in SGY and $-10^\circ$ in SGZ over the input ellipsoid. Finally, for the shifted case, we moved the center of the input ellipsoid by a vector $(-20.0, 80.0, -10.0)\> \mathrm{Mpc}/h$ in supergalactic coordinates.

\begin{table*}[!htbp]
\centering
\renewcommand{\arraystretch}{1.35}

\caption{
Input parameters adopted in the mock analysis. The ellipsoidal velocity field is injected only within the input geometry, while an independent dipolar component is assigned outside this region.
}

\label{tab:input_mock}

\begin{tabular}{|l|c|c|}
\hline\hline

Parameter
& Inside Input Geometry
& Outside Input Geometry \\[1.5mm]

\hline

Expansion $\theta$
& $10$ \footnotesize{km/s/Mpc}
& $-$  \\[1.5mm]

\hline

Dipole amplitude $|\mathbf{d}|$
& $100$ \footnotesize{km/s}
& $600$ \footnotesize{km/s}\\

Dipole direction (ra)
& $260^\circ$
& $100^\circ$ \\

Dipole direction (dec)
& $-20^\circ$
& $10^\circ$ \\[1.5mm]

\hline

Shear $\sigma_1$
& $-13$ \footnotesize{km/s/Mpc}
& $-$ \\

Shear $\sigma_2$
& $5$ \footnotesize{km/s/Mpc}
& $-$ \\

Shear $\sigma_3$
& $8$ \footnotesize{km/s/Mpc}
& $-$ \\[1.5mm]

\hline

Eigenvector $e_1$ (ra, dec)
& $(330.29^\circ,\, 13.51^\circ)$
& $-$ \\

Eigenvector $e_2$ (ra, dec)
& $(76.47^\circ,\, 49.24^\circ)$
& $-$ \\

Eigenvector $e_3$ (ra, dec)
& $(229.64^\circ,\, 37.5^\circ)$
& $-$ \\[1.5mm]

\hline

$H_0$
& \multicolumn{2}{c|}{$70$~\footnotesize{km/s/Mpc}} \\[1.5mm]

\hline

$\Ommo$
& \multicolumn{2}{c|}{$0.30$} \\[1.5mm]

\hline\hline

\end{tabular}
\end{table*}

\bibliographystyle{JHEP}
\bibliography{thisbib.bib}

\end{document}